\documentclass[11pt, letterpaper]{article}

\usepackage[margin=1in]{geometry}       % Standard 1-inch margins
\usepackage[onehalfspacing]{setspace}   % 1.5 spacing: The "Goldilocks" zone for reading
\usepackage{lineno}
\usepackage{soul}

\usepackage[utf8]{inputenc}
\usepackage[T1]{fontenc}
\usepackage{amsmath, amssymb, amsfonts}
\usepackage{bigints}
\usepackage{mathrsfs}
\usepackage{siunitx}
\DeclareSIUnit{\chain}{chains}
\DeclareSIUnit{\kBT}{\ensuremath{k_{\mathrm{B}}T}}

\usepackage{newtxtext}
\usepackage{newtxmath}

\usepackage{microtype} % Professional kerning/spacing

\usepackage{graphicx}
\usepackage{subcaption}
\usepackage[dvipsnames]{xcolor}
\usepackage{booktabs}  % Professional tables (\toprule, \bottomrule)
\usepackage{float}     % For [H] placement
\usepackage{siunitx}   % Materials science units

\usepackage[font=small, labelfont=bf, labelsep=period]{caption}

\usepackage[super, sort&compress, comma]{natbib}
\usepackage[colorlinks=true, linkcolor=NavyBlue, citecolor=NavyBlue, urlcolor=NavyBlue]{hyperref}
\usepackage{orcidlink} % Optional: Adds the little green ID icon

\usepackage{authblk}

\usepackage{titlesec}
\titleformat{\section}{\large\bfseries}{\thesection}{1em}{}
\titleformat{\subsection}{\normalsize\bfseries}{\thesubsection}{1em}{}

\title{\LARGE \textbf{Probing the many-body energy landscape of a soft glass with graph neural networks}}

\author[1]{Mehryar Jannesari Ghomsheh}
\author[1]{Donald L. Koch}
\author[1]{Sarah Hormozi\thanks{Corresponding author: \texttt{hormozi@cornell.edu}}}

\affil[1]{Robert Frederick Smith School of Chemical and Biomolecular Engineering, Cornell University, Ithaca, NY 14853, USA}

\date{}

\begin{document}

% \linenumbers

\maketitle

% ===========================================================
% ABSTRACT
% ===========================================================
\section*{Abstract} \label{sec:abstract}
\vspace{-6pt}
Many-body interactions govern the complex behavior of many amorphous materials, from metallic glasses to biological tissues, yet are often replaced by pairwise additive frameworks for computational efficiency. Here, we use classical density functional theory (DFT) to study a model soft glass of solvent-free polymer-grafted nanoparticles (PGNs), where the absence of solvent forces grafted chains to uniformly fill the interstitial space, generating strong angular-dependent many-body interactions between the cores. We show that NequIP, an equivariant message-passing graph neural network (GNN), predicts the equilibrium states with DFT-level accuracy, despite being trained only on high-energy, out-of-equilibrium configurations. Systematic analysis of GNN hyperparameters offers physical insights into the range, anisotropy, and effective body order of interactions. GNN-driven Monte Carlo simulations recover equilibrium structures in agreement with experiments and reveal locally favored icosahedral-like structures. Our results position machine-learned potentials as a tool for probing many-body glass physics beyond atomistic settings.

% ===========================================================
% MAIN TEXT
% ===========================================================

\section*{Introduction} \label{sec:introduction}
\vspace{-6pt}
Many-body interactions are fundamental to a wide range of disciplines, from social systems \cite{battiston2021physics} and ecological communities\cite{levine2017beyond} to colloidal science\cite{EricDufresneManybodyElectrostatics,boles2015many} and amorphous materials.\cite{cheng2009atomic, hohler2017many,bi2015density} The computational cost of accurately modeling higher-order interactions has limited many-body frameworks to small length and time scales.\cite{de2014relation} This presents a central challenge in characterizing the heterogeneous and slow nature of amorphous materials and has restricted our understanding of glass physics to pairwise-additive potentials that are fundamentally incapable of capturing higher-order interactions.\cite{tanaka2019revealing,li2024infinitely} Over the last decade, machine learning has shown great promise to model many-body interactions in a wide range of materials and molecules, presenting the accuracy of first-principles methods at a fraction of the cost.\cite{behler2007generalized, bartok2013representing, drautz2019atomic, schmidt2019recent,musil2021physics, schutt2018schnet, batatia2022mace, batzner2023, musaelian2023learning,cheng2024cartesian} However, whether machine learning approaches can successfully resolve the high-dimensional, rugged potential energy landscape (PEL) of many-body glassy systems remains an open question. \par 

The PEL of amorphous materials unravels many complex problems in glass physics.\cite{goldstein1969viscous,debenedetti2001supercooled} Relaxation dynamics can be modeled as exploration of the PEL through thermally activated transitions from one local minimum (inherent structure) to an adjacent one.\cite{nishikawa2022relaxation,berthier2023modern} Such local energy barriers are responsible for the slow dynamics of glasses.\cite{pica2024local} The PEL viewpoint explains the experimentally observed $\alpha$ (slow) and $\beta$ (fast) relaxations\cite{fan2014thermally,shiraishi2023johari} and enables predicting the location of future relaxation events as they strongly correlate with the energetic and spatial features of previous rearrangements.\cite{wang2020predicting,ji2025role} These valuable insights emerge from extensive exploration of the PEL, requiring millions of energy evaluations that are only tractable with pairwise-additive potentials.  \par
% The PEL viewpoint explains the experimentally observed $\alpha$ (slow) and $\beta$ (fast) relaxations respectively as escaping a meta-basin and transitioning between sub-basins within a deep meta-basin.\cite{fan2014thermally,shiraishi2023johari}

A new class of interaction potentials, called machine-learning interatomic potentials (MLIPs), has emerged to model many-body atomic interactions and accelerate ab-initio molecular dynamics simulations of large systems over long time scales. MLIPs are trained on quantum-mechanical calculations such as density functional theory (DFT) to predict the total potential energy of a system from the geometry of local atomic environments, capturing both distance- and angular-dependent many-body effects.\cite{schmidt2019recent,musil2021physics} Early approaches relied on hand-crafted descriptors of these environments,\cite{behler2007generalized,bartok2013representing} while more recent architectures employ message passing between atoms to learn many-body representations.\cite{schutt2018schnet,batatia2022mace,batzner2023} Several recent studies of metallic glasses have employed invariant, descriptor-based MLIPs trained on DFT data.\cite{xie2022neural,zhao2023development,wadowski2026efficient} Large barriers in the glassy PEL, combined with the high computational cost of modeling many-body interactions, restrict first-principles-driven sampling to a narrow and preparation-dependent portion of the landscape, while accurately learning the high-dimensional, rugged PEL demands a substantially large and configurationally diverse training dataset.\cite{bonfanti2025recent} MLIPs based on equivariant representations can potentially solve this problem by requiring significantly less training data while offering superior accuracy and generalization.\cite{batatia2022mace,batzner2023,musaelian2023learning} Such architectures naturally align with the underlying physical symmetries, thereby enhancing the interpretability of model hyperparameters.\cite{owen2024complexity} These advances have been developed within the atomistic setting, where the reference is quantum DFT. Soft glasses based on associative polymers\cite{gong2013modeling}, block co-polymers\cite{xi2019thermodynamics}, and various other types of polymer nanocomposites\cite{hall2010molecular} can be described by classical DFT and self-consistent field theories, yet MLIPs remain unexplored in this context.  \par

% \hl{Accurately learning the high-dimensional PEL of glasses demands a substantially large and configurationally diverse training dataset.}\cite{bonfanti2025recent} \hl{Recent strategies for metallic glasses include training on perturbed crystalline structures and high-temperature liquid states,}\cite{xie2022neural}\hl{as well as using surrogate models with single-point DFT corrections to access distinct supercooled states.}\cite{wadowski2026efficient}\hl{These approaches rely on invariant architectures, whereas MLIPs based on equivariant representations require significantly less training data while offering superior accuracy and generalization.} \cite{batatia2022mace,batzner2023,musaelian2023learning} \hl{Moreover, MLIPs remain unexplored in the context of soft glasses.}  \par

\begin{figure} [!ht]
    \centering
    \includegraphics[width=\textwidth]{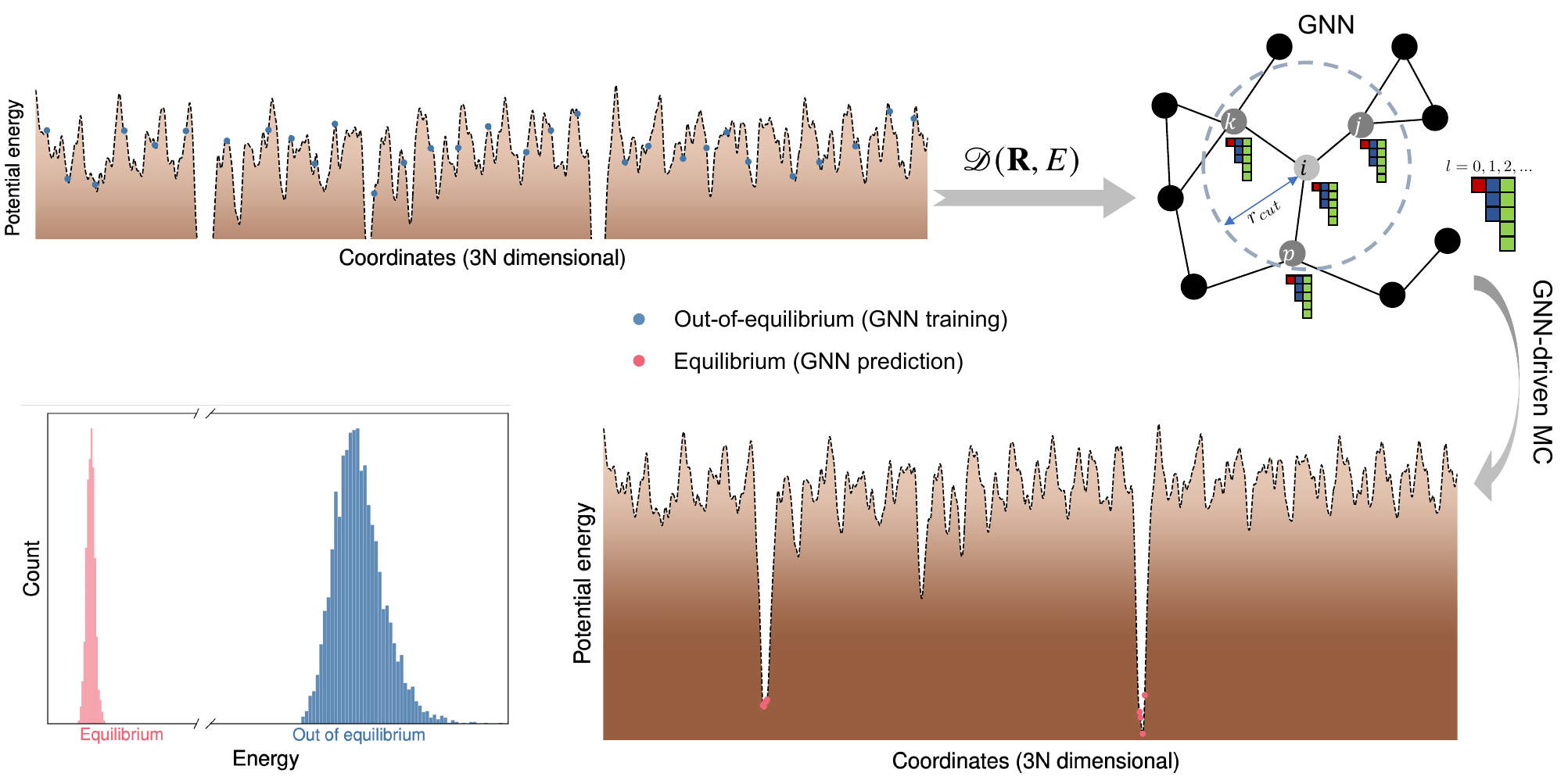}
    \caption{Overview of the computational framework. Out-of-equilibrium particle configurations are sampled via hard-sphere dynamics and their potential energies are evaluated with classical DFT, producing a dataset $\mathscr{D}(\mathbf{R}, E)$ that spans the high-energy region of the PEL (top left). The GNN is trained on this dataset. Each particle is represented as a node connected to its neighbors within a cutoff radius, $r_\mathrm{cut}$, and node features are updated using tensor representations of increasing angular resolution, $l=0, 1, 2, \dots$. The trained GNN drives MC simulations, enabling exploration of the deep minima of the PEL (bottom right) that are inaccessible during training. The GNN is validated by single-point classical DFT evaluations of the equilibrium configurations predicted by GNN-MC. The large difference between out-of-equilibrium and equilibrium energies (bottom left plot) demonstrates robust out-of-distribution generalization of the GNN.}
    \label{fig1}
\end{figure}

In this work, we train neural equivariant interatomic potential (NequIP)\cite{batzner2023} on classical DFT energies to learn the PEL of a model soft glass of solvent-free polymer-grafted nanoparticles (PGNs), see Fig. \ref{fig1}. NequIP is a graph neural network (GNN) where its internal features utilize not only scalars but also higher-order tensors that are equivariant with respect to the group of translations, rotations, and reflections in 3D space. This highly accurate and efficient modeling of angular many-body interactions makes NequIP potentially an ideal candidate for PGNs, in which the grafted polymer chains generate effective many-body interactions between particles.\cite{akcora2009anisotropic,kumar2013nanocomposites,zhang2025generalized} In PGNs embedded in a polymer matrix, previous studies show that steric repulsion from the grafted polymers creates angular-dependent three-body contributions irreproducible with pairwise-additive potentials.\cite{pryamtisyn2009modeling,tang2017anisotropic,zhou2023many} Polymeric interactions are enhanced in solvent-free conditions,\cite{chremos2011structural,wu2023implicit,moussavi2025nanoconfinement} suggesting stronger angular-dependent and higher-order many-body interactions, which are challenging to model. \par
The configurational and translational entropy of the grafted polymers controls the core configuration in solvent-free PGNs, resulting in a large free energy penalty for structural rearrangements, causing the particles to become caged by their neighbors.\cite{ghomsheh2026linking,agarwal2011strain,liu2019microscopic} Therefore, solvent-free PGNs can be viewed as jammed soft particles. Rheological experiments show that the system relaxes very slowly, exhibiting a soft glassy behavior, reminiscent of foams, emulsions, and tissues, over extended time periods.\cite{agarwal2010ages,agarwal2011strain,agarwal2012crowded,choudhury2015self,srivastava2015hyperdiffusive,wen2015dynamics,agrawal2016molecular,liu2019microscopic,parisi2021universal} This glassy behavior arises from polymer entropic constraints and can be observed at core volume fractions well below the colloidal glass transition, allowing the thermal fluctuations to relax the nanostructure towards equilibrium states over long, but experimentally accessible, time scales.\cite{srivastava2015hyperdiffusive,wen2015dynamics,liu2019microscopic} Our previously developed classical DFT predicts that the activation energy, and thus the macroscopic relaxation time, can be tuned by design parameters such as grafting density, polymer molecular weight, and core volume fraction, in agreement with experimental observations.\cite{ghomsheh2026linking} However, for computational tractability, these results were limited to face-centered-cubic (FCC) arrangement of cores. Here, we extend our framework to model the high-dimensional PEL. \par 
Probing equilibrium states requires long simulations with millions of energy evaluations which are computationally prohibitive with classical DFT. Instead, we perform GNN-driven Monte Carlo (MC) simulations using a NequIP model trained on classical DFT energies. The training configurations are generated by randomly sampling the PEL via hard-sphere dynamics (\hyperref[subsec:random_sampling]{Methods}), which yields a diverse set of configurations without energy bias but samples only out-of-equilibrium states (Fig. \ref{fig1}). Since the equilibrium states predicted by GNN-driven MC simulations are far outside the training distribution, we validate them with single-point classical DFT evaluations. We find that the GNN predicts equilibrium energies with DFT-level accuracy across all PGN design parameters studied, despite being trained only on high-energy, out-of-equilibrium structures. GNN-driven MC simulations predict the inter-particle spacing in agreement with small-angle X-ray scattering (SAXS) experiments and reveal locally favored structures. Furthermore, we extract fundamental physical insights into the nature of many-body interactions by analyzing the GNN's hyperparameters.

\section*{Many-body potential}
\begin{figure} [!ht]
    \centering
    \begin{subfigure}{0.3\textwidth}
        \centering
        \includegraphics[width=\textwidth]{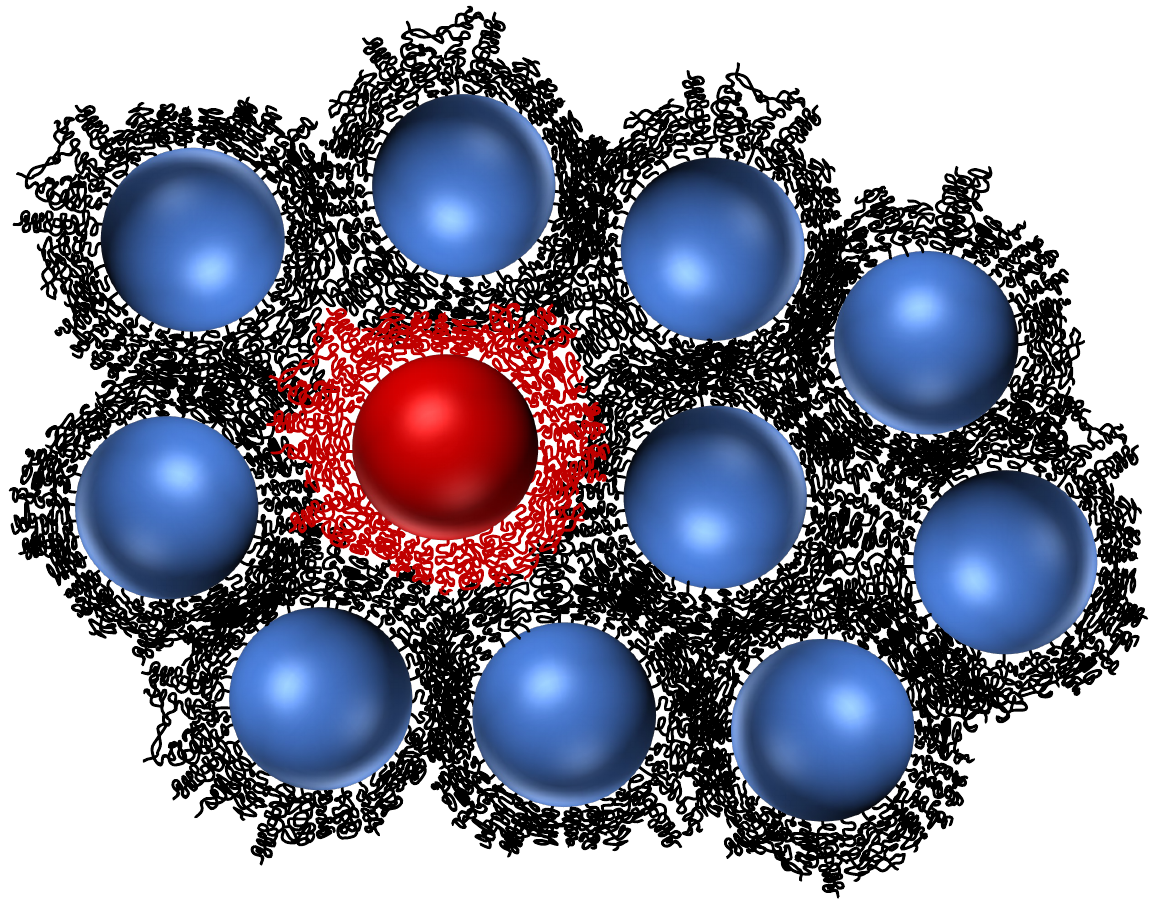}
        \caption{}
        \label{fig2a}
    \end{subfigure}
    \hfil
    \begin{subfigure}{0.3\textwidth}
        \centering
        \includegraphics[width=\textwidth]{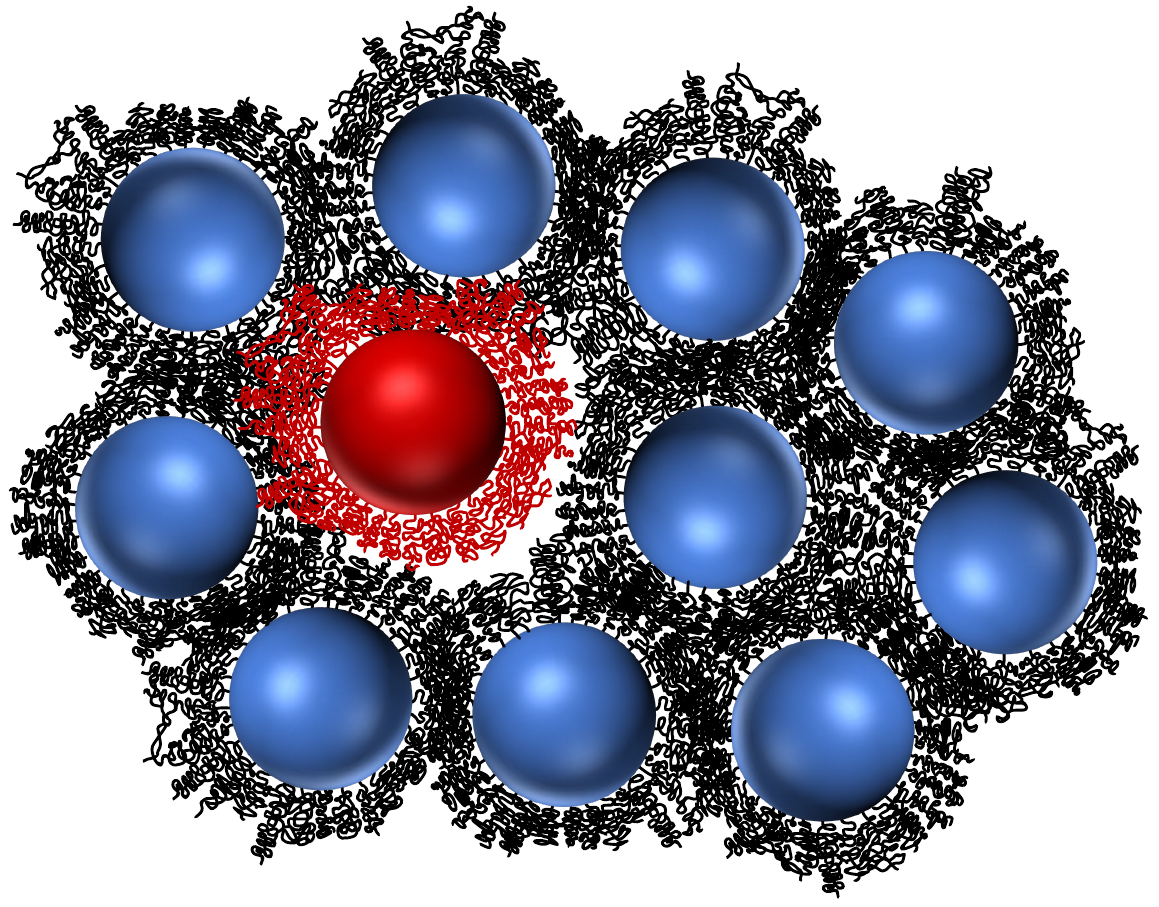}
        \caption{}
        \label{fig2b}
    \end{subfigure}
    \hfil
    \begin{subfigure}{0.3\textwidth}
        \centering
        \includegraphics[width=\textwidth]{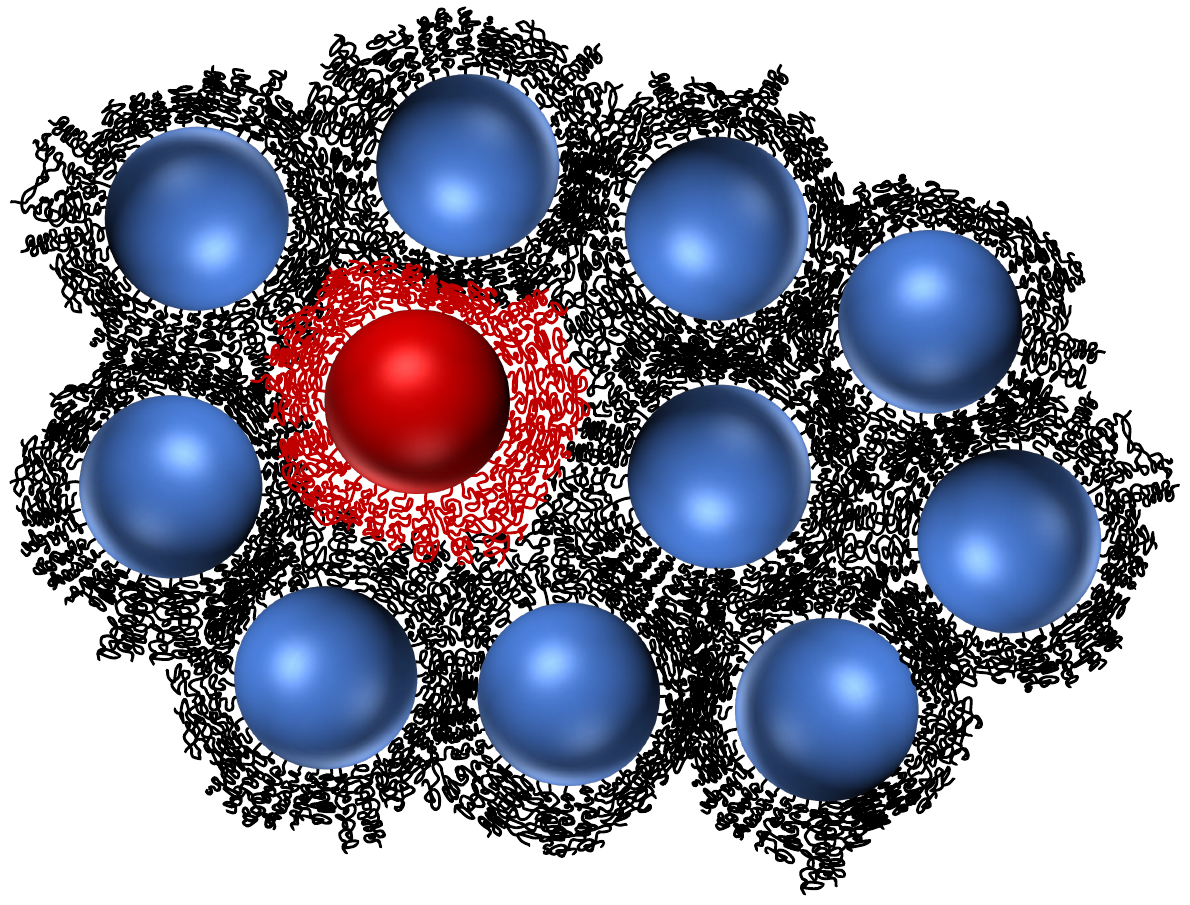}
        \caption{}
        \label{fig2c}
    \end{subfigure}
    \caption{Schematic of solvent-free PGNs. (a) Polymers uniformly fill the void space for a given configuration of particles. (b) Same configuration as (a) but with an arbitrary displacement of a random particle (shown in red). (c) Displacement of a single particle creates a cascade of polymer rearrangements to minimize the free energy by reaching a uniform number density.}
    \label{fig2}
\end{figure}

\vspace{-6pt}
Theoretical and experimental studies indicate that in solvent-free PGNs, the grafted polymers uniformly fill the interstitial space as an incompressible fluid as the particle's thermal energy is insufficient to compress the grafted polymers.\cite{yu2010structure,yu2014structure,srivastava2017self} A schematic of solvent-free PGNs is shown in Fig. \ref{fig2a}. Displacing a single particle creates local polymer density perturbations (Fig. \ref{fig2b}) that trigger a cascade of rearrangements involving polymers grafted to other core particles to establish a new thermodynamically favorable state of uniform number density (Fig. \ref{fig2c}). This change in the potential energy cannot be simplified to pairwise-additive contributions. We model polymer-mediated interactions between particles using classical DFT, drawing a conceptual analogy to how quantum DFT describes electron-mediated interactions between atoms. Relaxing the core configuration requires the collective movement of all grafted polymers, whereas individual polymer relaxation involves a single chain. We can therefore assume the polymers remain in equilibrium as they relax significantly faster than the cores.\cite{yu2010structure,chremos2011structure,yu2014structure} This separation of time scales plays a role similar to that of the Born–Oppenheimer approximation in quantum mechanics.\cite{martin2020electronic} Assuming the cores interact only through their polymers, for a given configuration of $N$ particles at positions $\mathbf{R}_1, \dots, \mathbf{R}_N$, the potential energy, $E(\mathbf{R}_1, \dots, \mathbf{R}_N)$, can be written based on the free energy of the grafted polymers (\hyperref[sec:methods]{Methods}).

\begin{figure} [!ht]
    \centering
    \begin{subfigure}{0.48\textwidth}
        \centering
        \includegraphics[width=\textwidth]{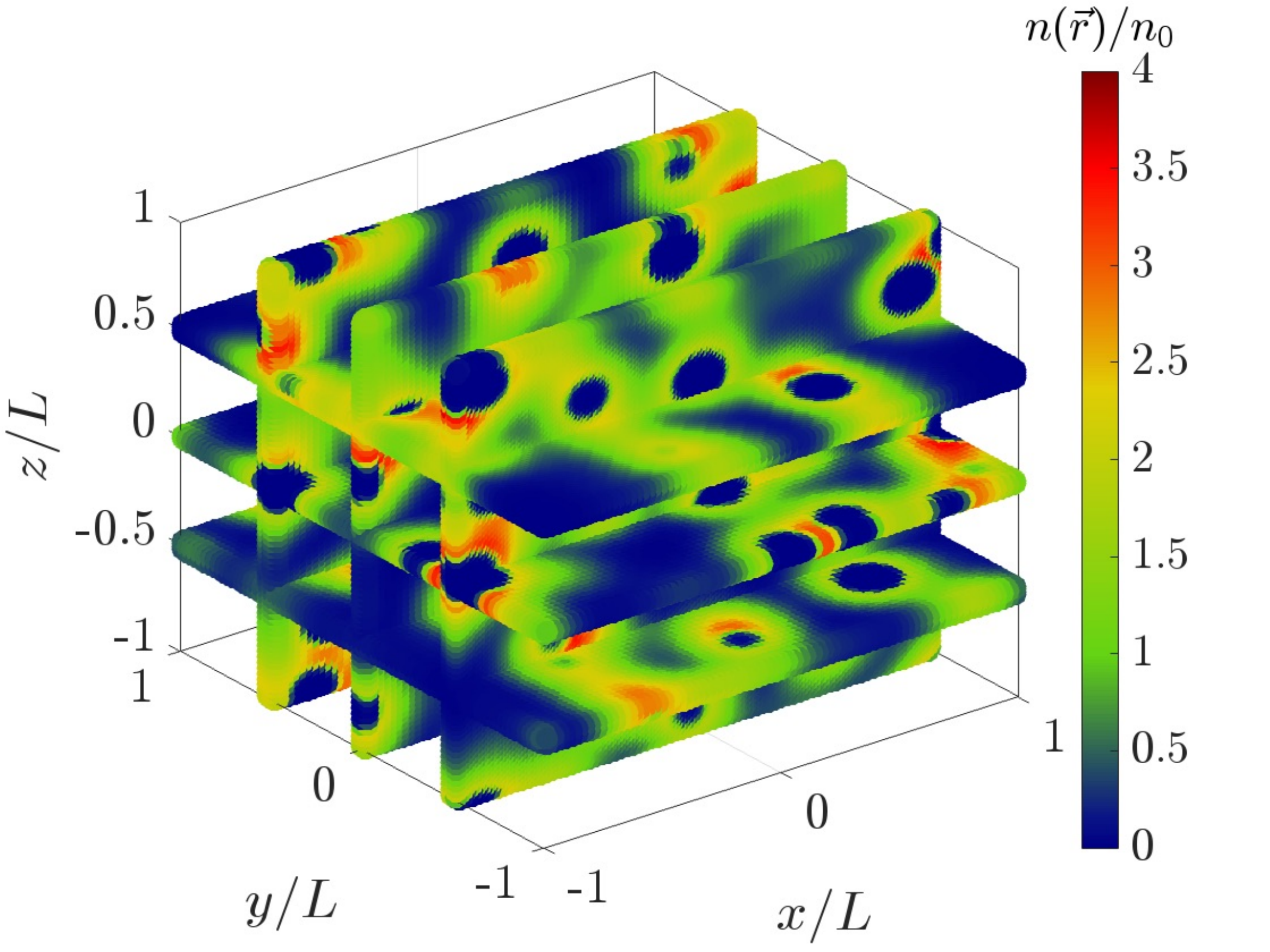}
        \caption{}
        \label{fig3a}
    \end{subfigure}
    \hfil
    \begin{subfigure}{0.48\textwidth}
        \centering
        \includegraphics[width=\textwidth]{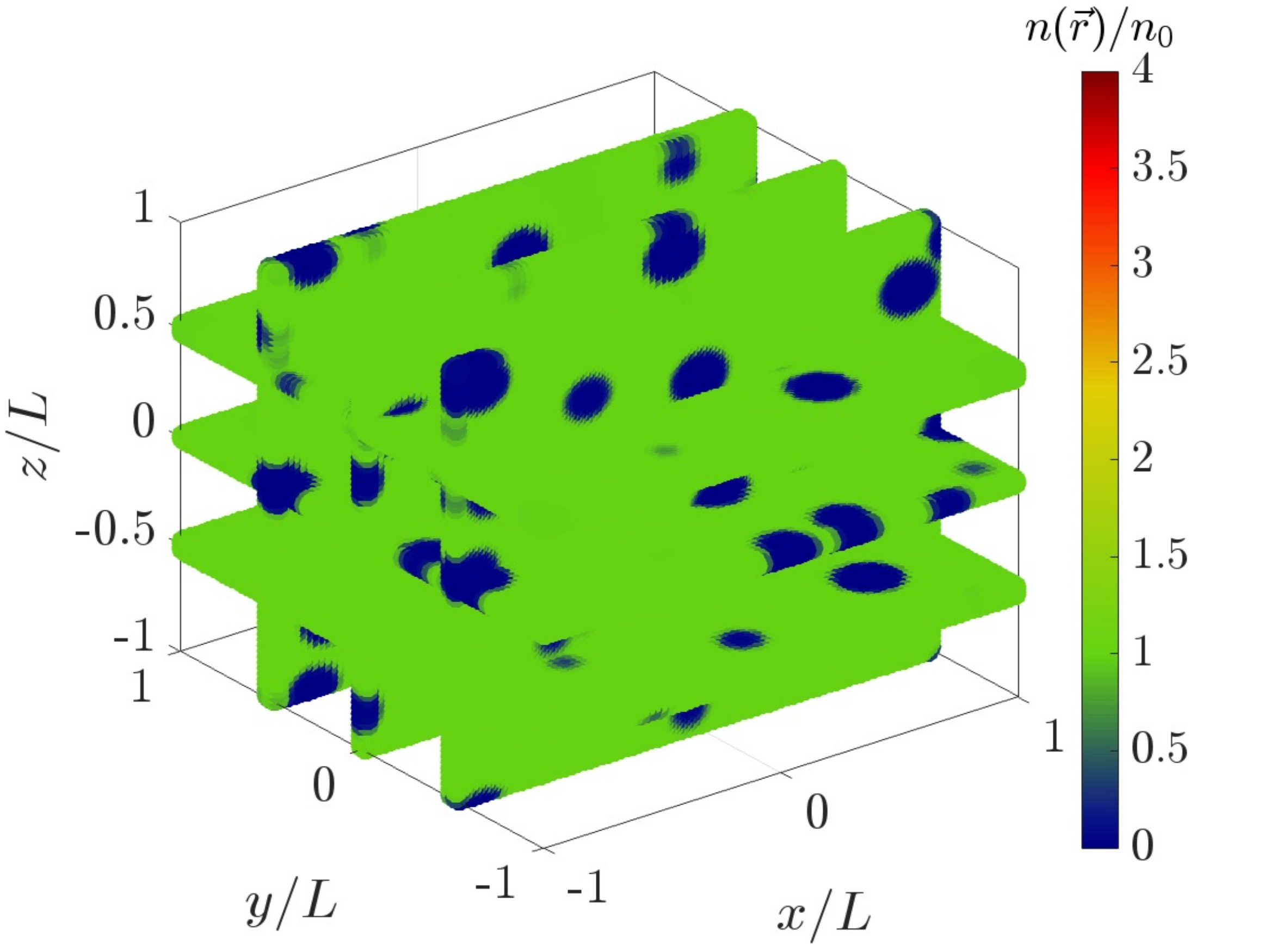}
        \caption{}
        \label{fig3b}
    \end{subfigure}
    \hfil
    \begin{subfigure}{0.48\textwidth}
        \centering
        \includegraphics[width=\textwidth]{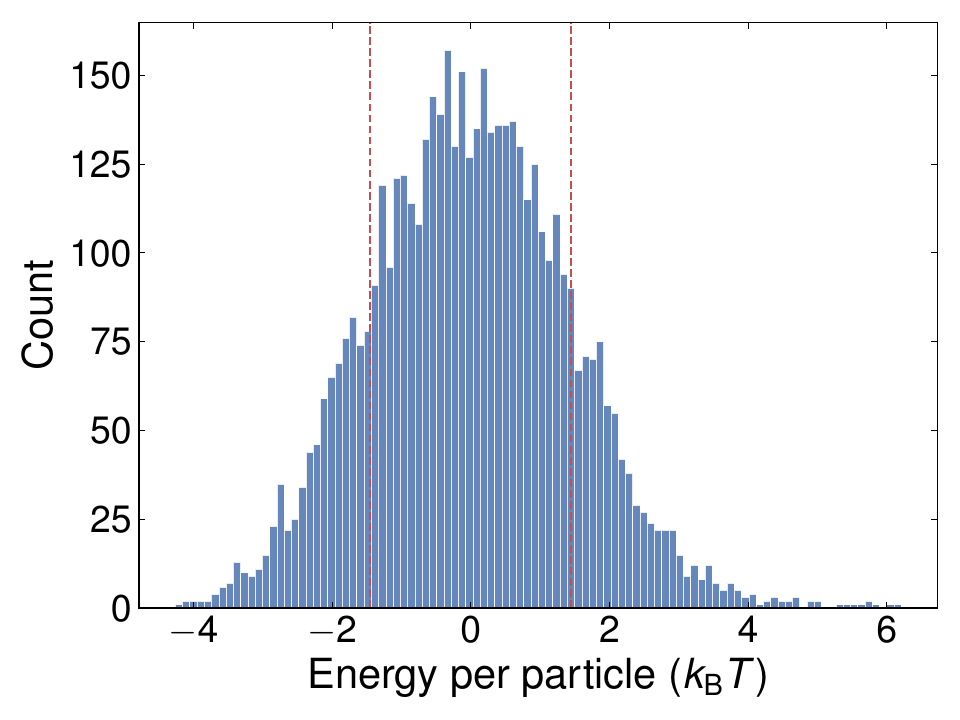}
        \caption{}
        \label{fig3c}
    \end{subfigure}
    \hfil
    \begin{subfigure}{0.48\textwidth}
        \centering
        \includegraphics[width=\textwidth]{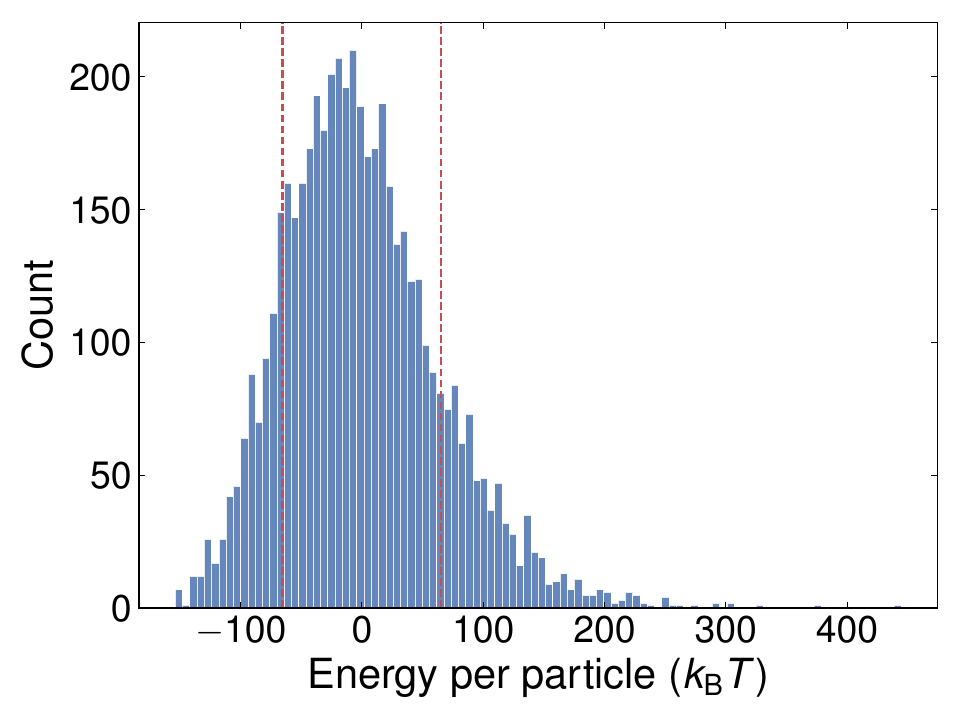}
        \caption{}
        \label{fig3d}
    \end{subfigure}
    \caption{Classical DFT results for \num{100} PGNs with a core volume fraction of $\phi_c = \num{0.1}$, polymer molecular weight of $M_w = \qty{5}{\kilo\dalton}$, grafting density of $\sigma_g = \qty{1.8}{\chain\per\nano\meter\squared}$, and core diameter of $d = \qty{10}{\nano\meter}$. The normalized number density of the grafted polymers is shown across multiple planes ($L$ denotes half of the box size) for (a) the system with excess implicit solvent, displaying significant density fluctuations, and (b) the solvent-free system, showing a uniform density distribution. The dark blue regions represent the cores. Per particle energy distributions, shifted by the mean, are shown for (c) the system with solvent and (d) the solvent-free system. The red vertical lines represent the standard deviation.}
    \label{fig3}
\end{figure}

The equilibrium polymer number density profiles are illustrated in Fig. \ref{fig3a} and Fig. \ref{fig3b} for a given configuration of PGNs with and without solvent, respectively. In the presence of solvent, chains localize near grafting sites to minimize their free energy, leaving solvent-filled gaps that effectively decouple the particles. Conversely, the strict incompressibility constraint in solvent-free conditions forces the polymers to stretch and uniformly fill regions even far away from their grafting location. This creates a highly connected network with significant polymer interpenetration, forms strong cages around the cores, and dramatically slows down the nanostructure relaxation. \par 
While a uniform density profile increases the translational entropy of the beads, it strongly suppresses the configurational entropy of the chains by restricting the number of accessible chain conformations. Since the loss of configurational entropy dominates, solvent-free PGNs have a much larger polymer free energy compared with PGNs with solvent.\cite{ghomsheh2026linking} This configurational entropic frustration is the molecular origin of slow dynamics and glassy behavior observed in solvent-free PGNs.\cite{agarwal2012crowded, yu2014structure} The manner in which the balance between translational and configurational entropy of polymer chains governs energy barrier formation is explained in depth in our previous work.\cite{ghomsheh2026linking} \par 
We initially probe the PEL by sampling high-dimensional configuration space of particles through $5000$ uncorrelated snapshots of equivalent hard spheres (\hyperref[subsec:random_sampling]{Methods}). The potential energy of each configuration is evaluated with classical DFT, and the resulting energy distributions are demonstrated in Fig. \ref{fig3c} and Fig. \ref{fig3d} for PGNs with and without solvent, respectively. With solvent, the energy distribution is nearly Gaussian with a mild right tail due to  polymer-core excluded volume interactions. A pronounced long tail appears for high energies in the absence of solvent, corresponding to the increased cost of filling large void regions. The solvent-free system exhibits a substantially larger energy variance, reflecting greater heterogeneity across the PEL. This necessary, though not sufficient, condition for a rugged landscape with deep minima is absent from the system with solvent. Increasing the core volume fraction at fixed polymer molecular weight or increasing the molecular weight at fixed core volume fraction reduces the heterogeneity in the PEL (Supplementary Fig. \ref{sup-fig-energydist}), owing to mitigated entropic frustration as the grafted chains are required to fill a smaller or more accessible void volume per core.\cite{ghomsheh2026linking} We examine the effects of landscape heterogeneity on out-of-distribution generalization accuracy of the GNN. For the remainder of this study, we only focus on solvent-free conditions.  \par 

\section*{Graph neural network}
\begin{figure} [!ht]
    \centering
    \begin{subfigure}{0.48\textwidth}
        \centering
        \includegraphics[width=\textwidth]{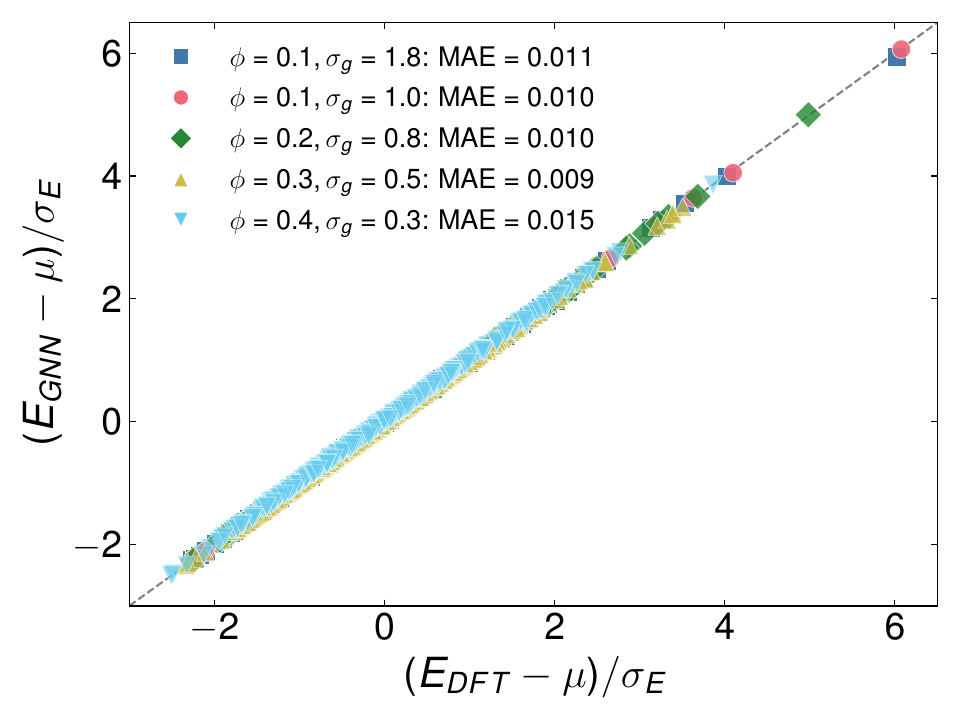}
        \caption{}
        \label{fig4a}
    \end{subfigure}
    \hfil
    \begin{subfigure}{0.48\textwidth}
        \centering
        \includegraphics[width=\textwidth]{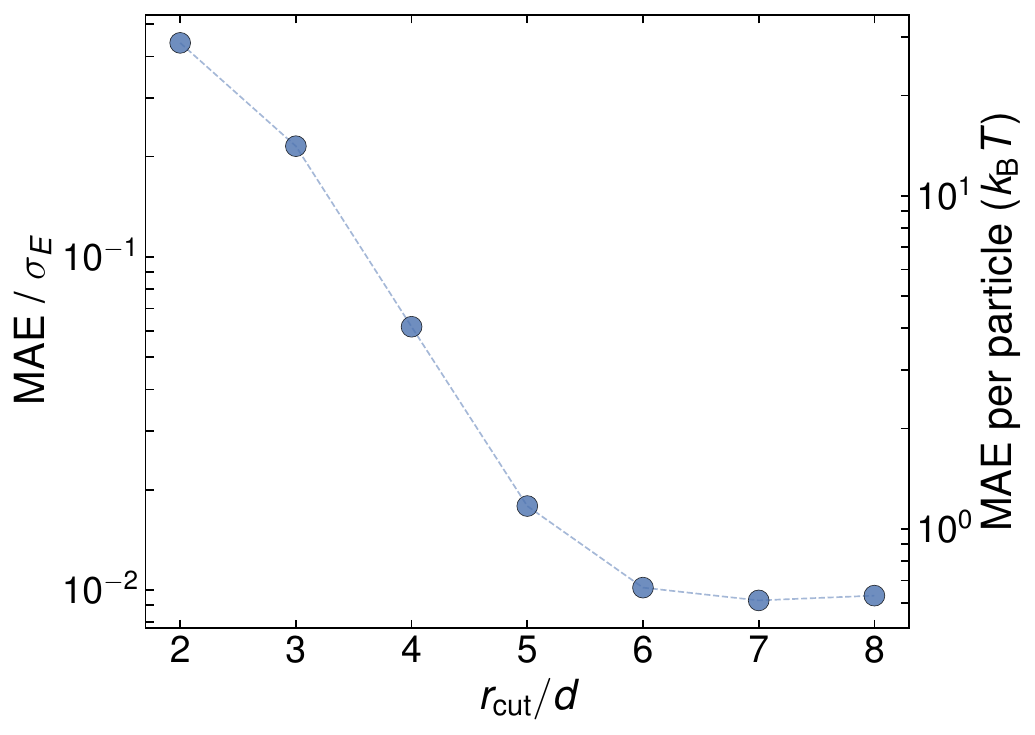}
        \caption{}
        \label{fig4b}
    \end{subfigure}
    \hfil
    \begin{subfigure}{0.48\textwidth}
        \centering
        \includegraphics[width=\textwidth]{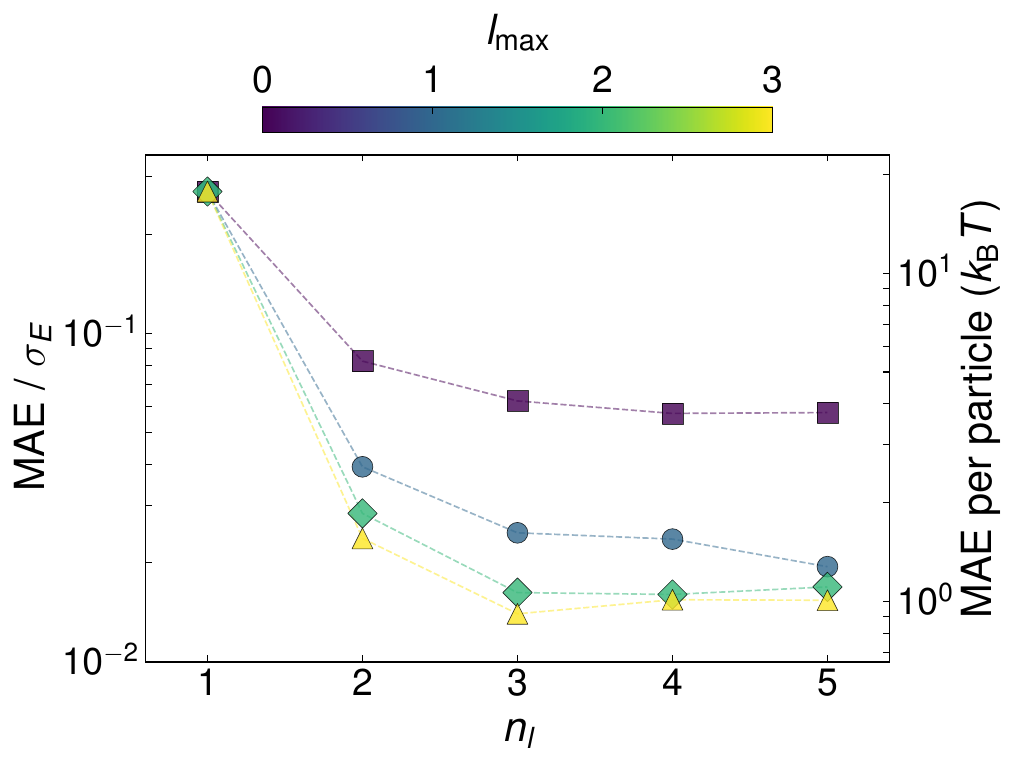}
        \caption{}
        \label{fig4c}
    \end{subfigure}
    \hfil
    \begin{subfigure}{0.48\textwidth}
        \centering
        \includegraphics[width=\textwidth]{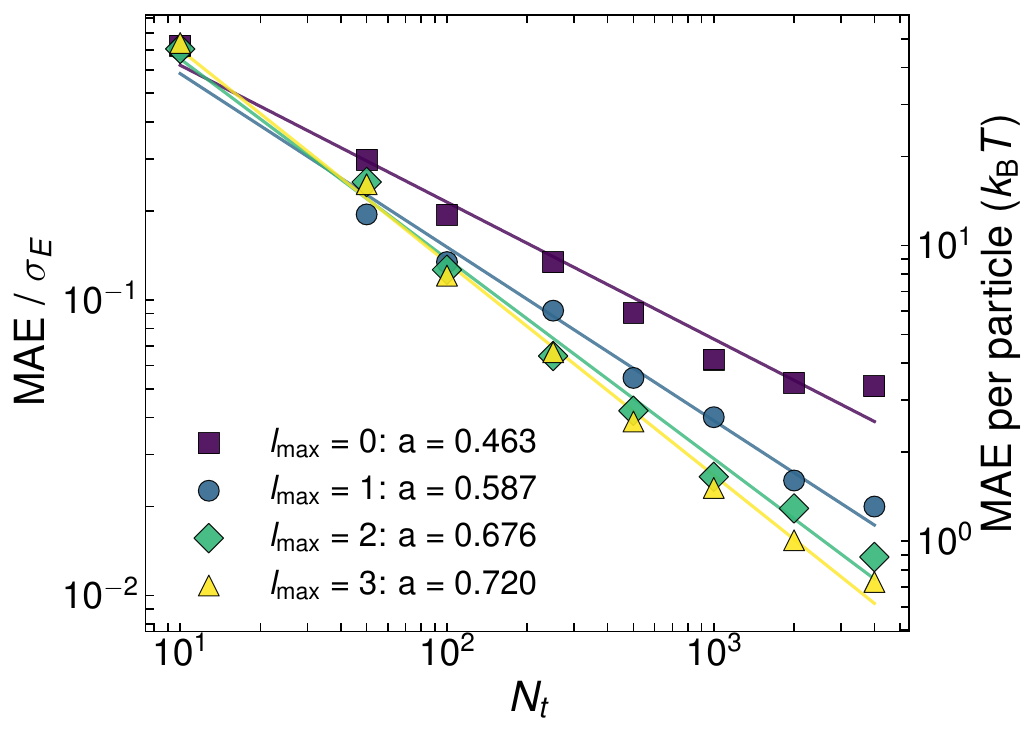}
        \caption{}
        \label{fig4d}
    \end{subfigure}
    \caption{(a) Parity plot comparing GNN predictions to classical DFT energies for solvent-free PGNs across varying design parameters. The energies are shifted by the mean value, $\mu$, and scaled by the standard deviation, $\sigma_E$, of each sample. All samples have a polymer molecular weight of $M_w= \qty{5}{\kilo\dalton}$, except for the sample with a grafting density of $\sigma_g = \qty{1}{\chain\per\nano\meter\squared}$ which has $M_w = \qty{9}{\kilo\dalton}$. (b) The energy MAE as a function of the normalized cutoff radius. In (b)-(d), the left y-axis corresponds to energy MAE of validation dataset normalized by the standard deviation of energy, and the right y-axis shows MAE per particle. (c) The energy MAE as a function of number of message-passing layers for different angular resolutions. (d) Learning curves illustrating energy MAE as a function of training set size for different angular resolutions. Tests in (b)-(d) correspond to the sample with $\sigma_g=\qty{1.8}{\chain\per\nano\meter\squared}$.}
    \label{fig4}
\end{figure}

\vspace{-6pt}
The analysis above provides molecular-level insights into the PEL of solvent-free PGNs and soft glasses in general. However, our sampling primarily accesses high-energy configurations and does not capture local minima that govern equilibrium and near-equilibrium behavior. We train NequIP on this dataset to develop a computationally efficient inter-particle potential and accelerate sampling of equilibrium states. Fig. \ref{fig4a} compares the GNN predictions with the ground-truth classical DFT energies for the test dataset. Since only energy differences in our classical DFT are relevant, the energies are shifted by the mean and scaled by the standard deviation, enabling comparison across samples. Across the full range of PGN design parameters studied, the network predicts classical DFT energies with a mean absolute error (MAE) of around 0.01 $\sigma_E$, where $\sigma_E$ is the standard deviation of energy. This corresponds to per particle MAEs below the thermal energy scale, $\unit{\kBT}$, relevant for structural organization (see Supplementary Fig. \ref{sup-fig-gnn} for per-particle energies). The training and validation errors are reported in Supplementary Table \ref{sup-table-Tr-Val-Test-Error}. These results correspond to the optimal set of NequIP's hyperparameters (see \hyperref[subsec:nequip]{Methods} section for the definition of these hyperparameters and the optimal values chosen in this study). \par
A striking feature of the optimal models is the large cutoff radius required for accurate predictions. The MAE drops by more than an order of magnitude as $r_{\mathrm{cut}}/d$ increases from $2$ to $6$ before saturating (Fig. \ref{fig4b}). This implies that the effective interaction range extends well beyond the immediate particle neighborhood. This is not a signature of long-range pair interactions, as direct PGN interactions require corona overlap and are therefore inherently short-ranged, but rather of many-body correlations that extend the effective interaction range. To disentangle these many-body contributions, we vary the number of message-passing layers, $n_l$, while holding other hyperparameters fixed. For this test, we omit nonlinear activations between message-passing layers to prevent the introduction of implicit higher-order correlations.\cite{batatia2025design} Accuracy improves monotonically with $n_l$ up to three layers across all angular resolutions, $l_{\mathrm{max}}$, tested (Fig. \ref{fig4c}). No statistically significant gains are observed beyond $n_l=3$. Since each message-passing layer increments the correlation order by one in the linear regime, this convergence pinpoints the crucial effects of four-body correlations in energy predictions. In contrast, for PGNs in solvent, three-body interactions are generally considered sufficient.\cite{tang2017anisotropic,zhou2023many} We attribute these higher-order interactions to the cooperation of grafted polymers to uniformly fill the interstitial space. These higher-order correlations are strongly directional, a consequence of anisotropic local particle environments that impose preferential directions on how grafted polymers configure themselves to fill the interstitial space. For a fixed $n_l$ (except for $n_l=1$ which represents a pairwise network), the MAE decreases sharply upon switching from invariant ($l_{\mathrm{max}} = 0$) to equivariant ($l_{\mathrm{max}} \geq 1$) networks. \par 
The data-efficiency of different network architectures further corroborates this picture. Generalization error follows the typical power-law scaling $\epsilon \thicksim N_{t}^{-a}$ with the number of training examples $N_t$ (Fig. \ref{fig4d}). Equivariant networks consistently outperform invariant ones and, crucially, learn faster, reflected in larger exponents $a$. Within equivariant networks, higher angular resolution continues to improve data efficiency, an effect we attribute to capturing intricate angular many-body interactions that are revealed as new data become available. An analogous acceleration in learning is observed when the number of message-passing layers is increased instead (see Supplementary Fig. \ref{sup-fig-lc-nl}), suggesting that matching the network architecture to the physics accelerates learning. \par 

\section*{GNN-driven Monte Carlo simulations}
\begin{figure} [!ht]
    \centering
    \begin{subfigure}{0.48\textwidth}
        \centering
        \includegraphics[width=\textwidth]{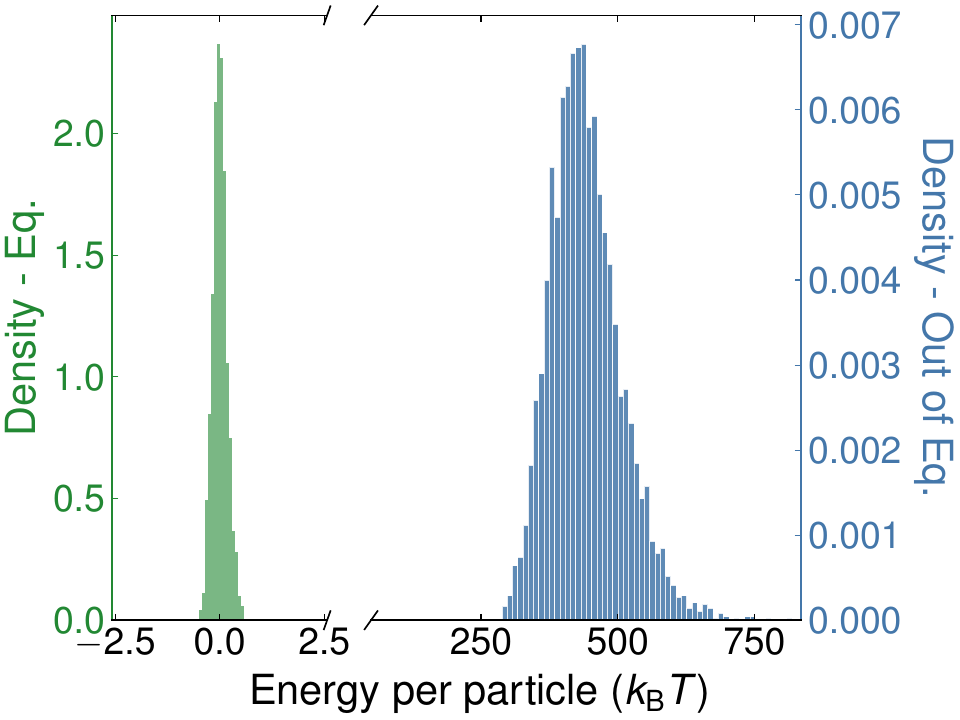}
        \caption{}
        \label{fig5a}
    \end{subfigure}
    \hfil
    \begin{subfigure}{0.48\textwidth}
        \centering
        \includegraphics[width=\textwidth]{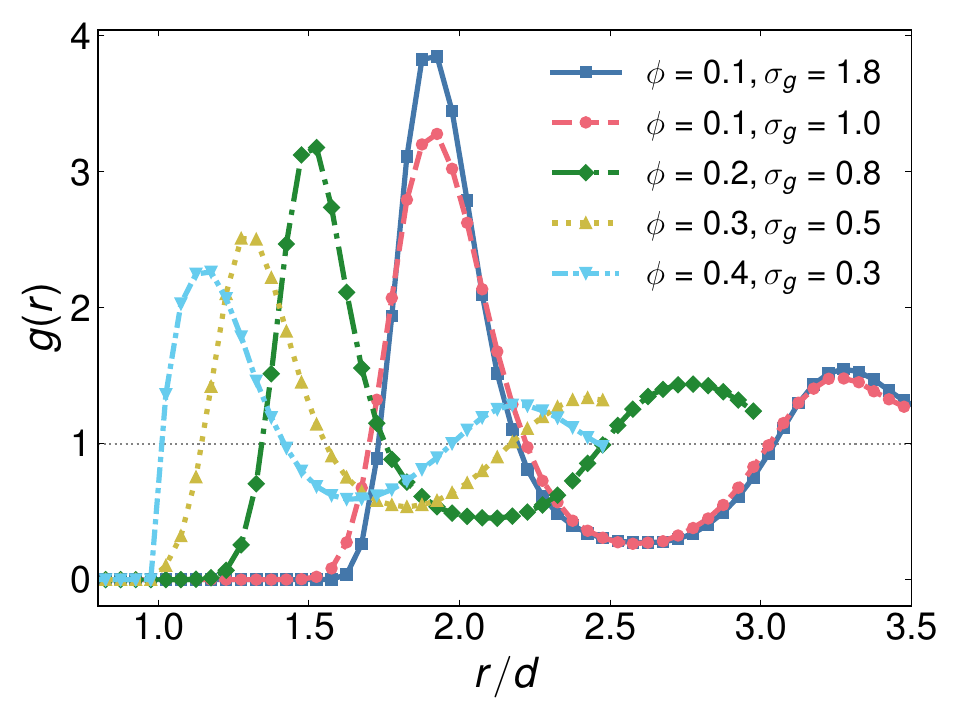}
        \caption{}
        \label{fig5b}
    \end{subfigure}
    \hfil
    \begin{subfigure}{0.48\textwidth}
        \centering
        \includegraphics[width=\textwidth]{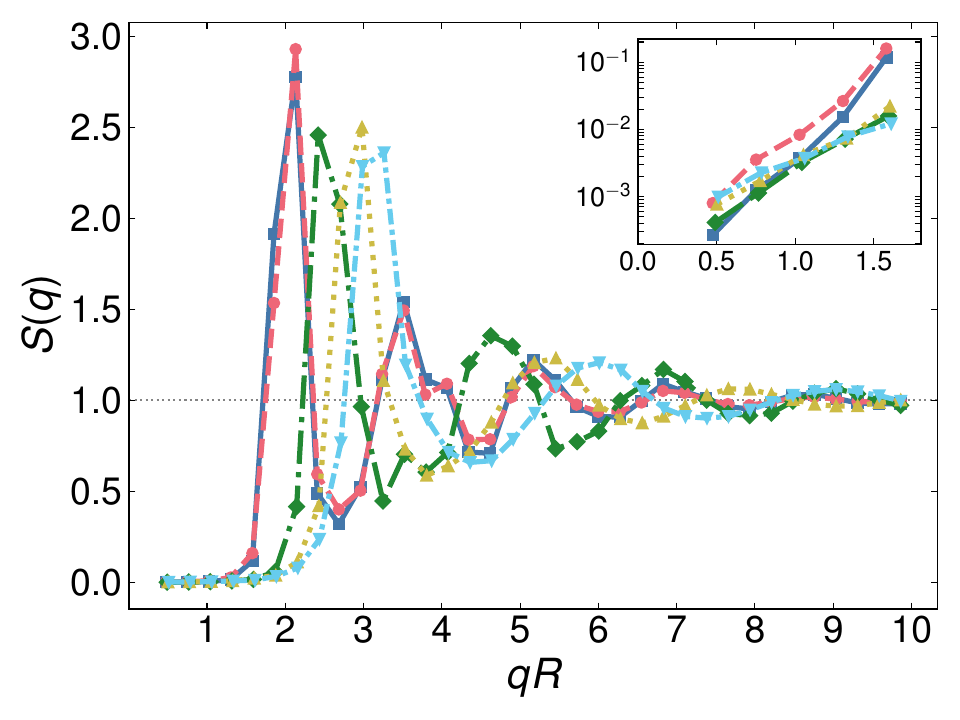}
        \caption{}
        \label{fig5c}
    \end{subfigure}
    \hfil
    \begin{subfigure}{0.48\textwidth}
        \centering
        \includegraphics[width=\textwidth]{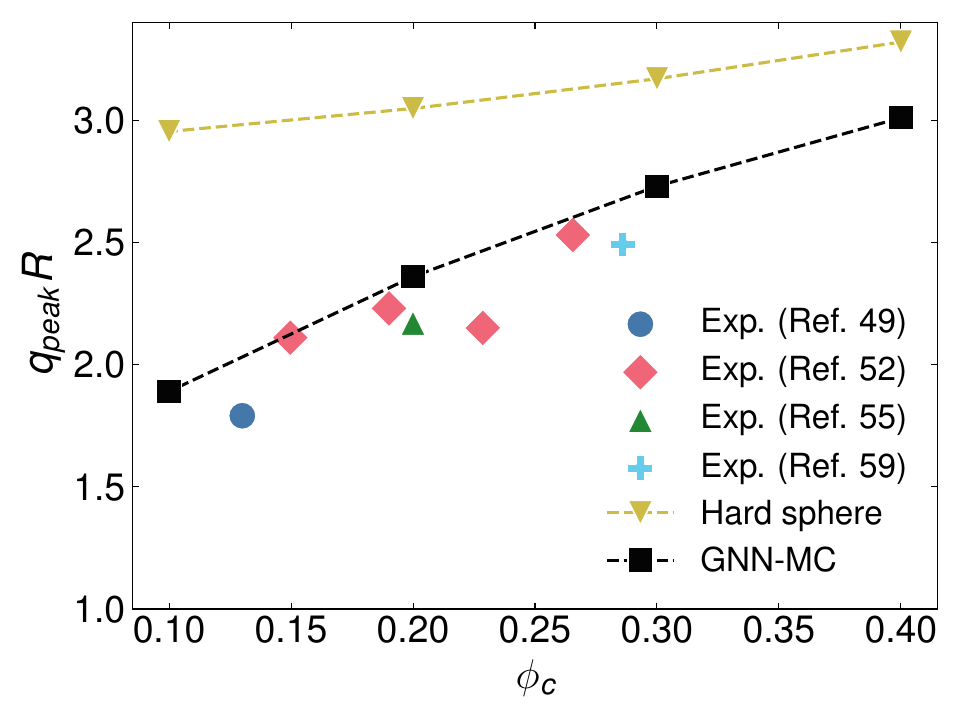}
        \caption{}
        \label{fig5d}
    \end{subfigure}
    \caption{(a) Per particle energy distribution for out-of-equilibrium configurations (shown in blue, sampled by hard-sphere dynamics and used for training GNN) and equilibrium configurations (shown in green) evaluated by classical DFT for the sample with $\phi_c=\num{0.1}$ and $\sigma_g = \qty{1.8}{\chain\per\nano\meter\squared}$. The equilibrium configurations are obtained via GNN-driven MC simulations. Both distributions are shifted by the mean energy of equilibrium configurations. (b) Equilibrium pair distribution function and (c) static structure factor for solvent-free PGNs with different design parameters. The inset shows the same data in the small-$qR$ regime on a log–log scale, where $R$ is the particle radius. (d) The location of the first peak of the static structure factor predicted by GNN-driven MC simulations, compared with experimental measurements for solvent-free PGNs, across a range of core volume fractions. All the samples have a polymer molecular weight of $M_w = \qty{5}{\kilo\dalton}$. The prediction for a hard-sphere liquid at the same $\phi_c$ is shown for comparison.}
    \label{fig5}
\end{figure}

\vspace{-6pt}
While retaining the DFT-level accuracy, the GNN reduces the cost of energy evaluations by four orders of magnitude, enabling long MC simulations to reach equilibrium. Fig. \ref{fig5a} shows the energy distribution of out-of-equilibrium configurations (used for training the GNN) and the equilibrium energies predicted by GNN-driven MC simulations (see \hyperref[subsec:mc]{Methods} section for details on MC simulation). The equilibrium distribution lies well outside the training range. To validate out-of-distribution accuracy, we passed the GNN-predicted equilibrium configurations to the classical DFT. GNN and classical DFT equilibrium energy distributions differ by a shift of up to $\unit{\kBT}$ per particle depending on the sample (Supplementary Table \ref{sup-table-DFT-GNN-Error} and Supplementary Fig. \ref{eq-energies-dft-vs-neq}). This error in the mean does not affect MC sampling 
since only energy differences enter the Metropolis criterion. The GNN predicts energy differences among equilibrium configurations with a per particle $\text{MAE}_{\Delta E}$ of 
$\qty{0.15}{\kBT}$, more than three 
orders of magnitude smaller than the separation between the training and equilibrium 
energy distributions for the most glassy sample studied (Supplementary Table \ref{sup-table-DFT-GNN-Error}). The error decreases for less glassy landscapes as the training and equilibrium energy distributions approach one another (Supplementary Table \ref{sup-table-DFT-GNN-Error} and Supplementary Fig. \ref{eq-energies-dft-vs-neq}). \par
The equilibrium pair distribution function (Fig. \ref{fig5b}) exhibits expected trends: decreasing core volume fraction shifts the first peak to larger separations and strongly suppresses particle contacts. For a fixed $\phi_c$, reducing the grafting density by increasing polymer molecular weight has little effect on peak position but reduces the peak height. This arises from a milder space-filling penalty imposed on longer chains, permitting greater thermal fluctuations of the cores. At low $\phi_c$, the polymers are significantly stretched while at high $\phi_c$, the polymers are compressed in some regions. The equilibrium structures are substantially different from the out-of-equilibrium structures used for training (Supplementary Fig. \ref{HS-structure}). Equilibrium Voronoi volume distribution is markedly narrow to accommodate a uniform polymer stretch. By contrast, out-of-equilibrium configurations display a broad Voronoi volume distribution (Supplementary Fig. \ref{voronoi}), particularly at low $\phi_c$.  \par 
A distinctive feature of solvent-free PGNs is evidenced in the static structure factor (Fig. \ref{fig5c}): long-range density fluctuations are strongly suppressed across all design parameters studied. In the monodisperse limit, each particle must exclude exactly one neighbor, leading to an incompressible single-component fluid with $S(q \rightarrow 0) \rightarrow 0$.\cite{yu2010structure,chremos2011structure} Polydispersity in experiments raises $S(0)$ slightly above zero, yet it remains well below that of a hard-sphere suspension at the same core volume fraction,\cite{yu2014structure,srivastava2017self} confirming that the equilibrium structures are disordered hyperuniform.\cite{srivastava2017self,chremos2017particle} The out-of-equilibrium snapshots do not show hyperuniformity as they follow the Percus–Yevick approximation for hard-sphere liquids (Supplementary Fig. \ref{Sq-hs}).  \par
The location of the first peak of $S(q)$ predicted by the GNN-driven MC simulations agrees well with SAXS measurements across all core volume fractions studied (Fig. \ref{fig5d}). This provides a measure of the characteristic inter-particle spacing at equilibrium. The deviation from the hard-sphere prediction shows that inter-particle distances are governed by the polymer-mediated inter-particle potential rather than core packing alone. The small discrepancies between our predictions and experiments can be mainly attributed to unavoidable polydispersity in experiments (in particle radius, polymer molecular weight, and grafting density) as well as simplifying assumptions of our classical DFT.

% The ratio $h = S(0)/S_p$ where $S_p$ is the value of $S(q)$ at the first peak, measures the degree of hyperuniformity. For a perfect hyperuniform material $h=0$, whereas for $h < 10^{-3}$, the material is considered effectively hyperuniform.\cite{torquato2018hyperuniform, kim2018effect, chremos2018hidden} The structure factor does not seem to reach a plateau across all the samples studied (Fig. \ref{fig5c}), suggesting $h \rightarrow 0$ as expected for monodisperse solvent-free PGNs. The criterion $h < 10^{-3}$ is already satisfied for the smallest wavenumber calculated in this study.

\subsection*{Locally favored structures}
\begin{figure} [!ht]
    \centering
    \begin{subfigure}{0.32\textwidth}
        \centering
        \includegraphics[width=\textwidth]{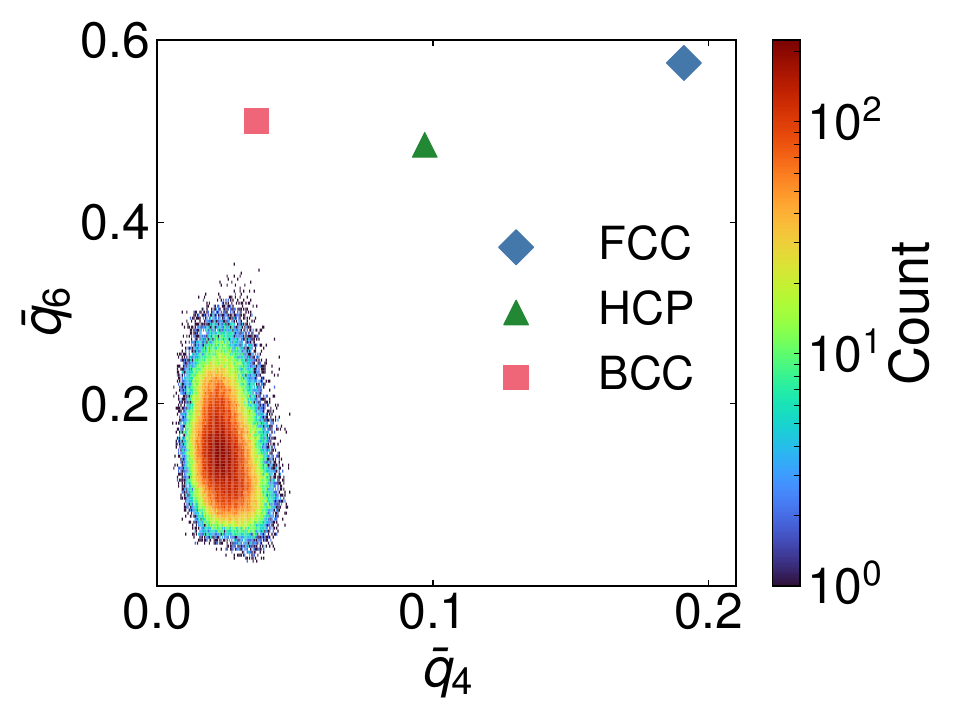}
        \caption{}
        \label{fig6a}
    \end{subfigure}
    \hfil
    \begin{subfigure}{0.32\textwidth}
        \centering
        \includegraphics[width=\textwidth]{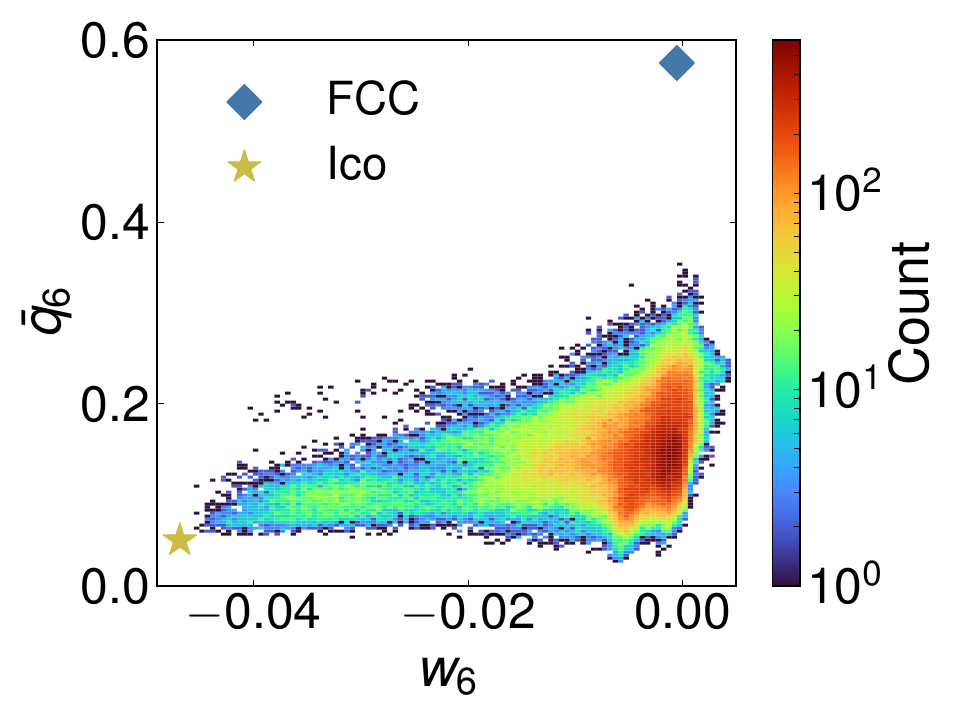}
        \caption{}
        \label{fig6b}
    \end{subfigure}
    \hfil
    \begin{subfigure}{0.32\textwidth}
        \centering
        \includegraphics[width=\textwidth]{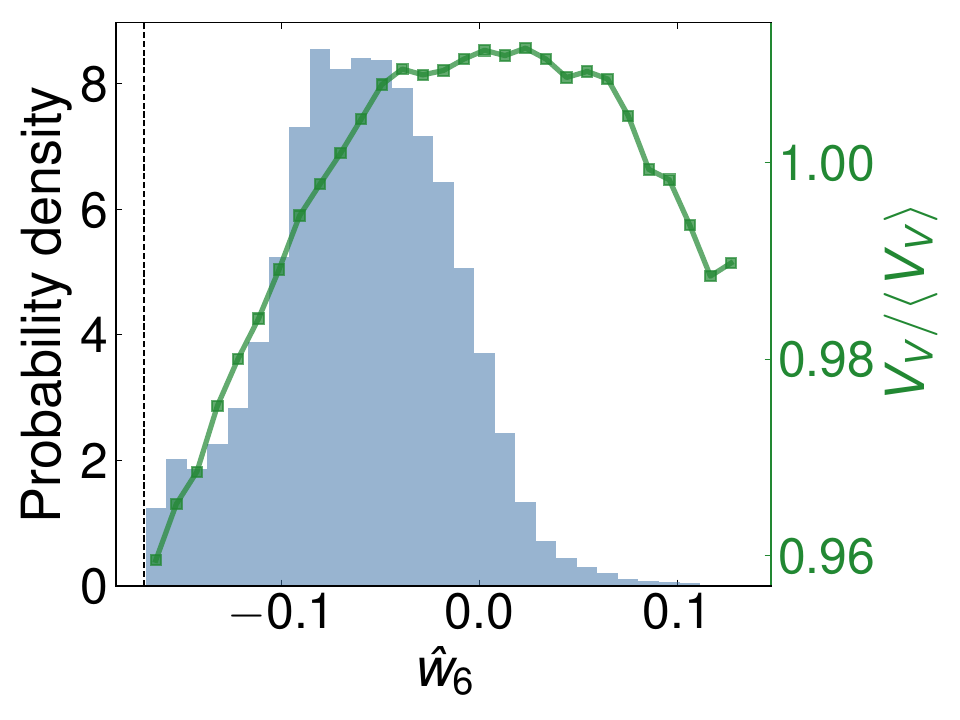}
        \caption{}
        \label{fig6c}
    \end{subfigure}
    \caption{Distribution of BOO parameters characterizing local particle configurations in solvent-free PGNs with $\phi_c = \num{0.1}$, $\sigma_g = \qty{1.8}{\chain\per\nano\meter\squared}$, and $M_w = \qty{5}{\kilo\dalton}$. The characteristic BOO values associated with perfect reference structures are marked on each map. (a) The $(\bar{q}_4, \bar{q}_6)$ plane and (b) the $(w_6, \bar{q}_6)$ plane. (c) Distribution of normalized $\hat{w}_6$ and the Voronoi volumes of the structures in each $\hat{w}_6$ bin normalized by the average Voronoi volume. The dashed vertical line represents $\hat{w}_6$ of a perfect icosahedron.}
    \label{fig6}
\end{figure}

\vspace{-6pt}
We characterize local particle environments using Steinhardt's bond orientational order (BOO) parameters\cite{steinhardt1983bond} and their variants\cite{lechner2008accurate} (\hyperref[subsec:structure]{Methods}). We observe no trace of crystal-like structures (Fig. \ref{fig6a}). This conclusion is further corroborated by MC simulations initialized from a perfect FCC lattice, which relax to the same disordered equilibrium structure as runs started from disordered configurations (Supplementary Fig. \ref{fcc-vs-hs}). \par 
Within this disordered landscape, the $(w_6, \bar{q}_6)$ map reveals a clear signature of local icosahedral-like ordering at low $\phi_c$ and high grafting densities where the system shows a glassy behavior (Fig. \ref{fig6b}). The five-fold symmetry of icosahedra is incompatible with long-range periodicity, and thus limits the number of icosahedral-like structures. Icosahedral ordering is regarded as reference structure in amorphous materials particularly in dense hard-sphere packing and hard-sphere glasses,\cite{van1995real,leocmach2012roles} soft jammed materials,\cite{vasisht2020emergence,vinutha2024memory} and dense metallic liquids and glasses\cite{reichert2000observation,ding2017computational} and is often linked to maximizing local density of packing and enhancing attractive interactions.\cite{frank1952supercooling,de2015entropy} We find strong correlations between icosahedral-like structures and Voronoi volumes (Fig. \ref{fig6c}). Such locally dense structures are only present at equilibrium (Supplementary Fig. \ref{w6-hs-vs-eq}) as they provide a thermodynamically favorable state for grafted polymers. The dodecahedral Voronoi volume associated with icosahedral particle configurations is geometrically close to a sphere,\cite{teich2016clusters} which is the most favorable shape for grafted polymers since it leads to uniform polymer stretches. \par

\section*{Discussion}
\vspace{-6pt}
We demonstrated that many-body interactions in amorphous materials can be accurately captured by NequIP, an equivariant GNN. As a model soft glass, we studied solvent-free PGNs in which grafted polymers uniformly fill the interstitial space giving rise to angular-dependent many-body interactions between the particles. We found that the range of energies in the PEL can be tuned through design parameters such as core volume fraction, grafting density, and polymer molecular weight. A detailed analysis of the GNN's hyperparameters revealed strongly directional many-body interactions, with correlations up to fourth order, that extend well beyond the nearest-neighbor shell. GNN-driven MC simulations reproduced previously reported structural features, including hyperuniformity, suppression of particle contacts, and enhanced nearest-neighbor distances, and additionally revealed local icosahedral structures at equilibrium. \par 
Previous studies showed that low core volume fraction and high grafting densities produce large activation energies\cite{ghomsheh2026linking} and strong cages around the particles\cite{srivastava2017self} due to greater degrees of polymer entropic frustration. We found that this leads to a highly heterogeneous PEL and the appearance of local icosahedral ordering at equilibrium. Establishing a quantitative connection between PEL topology, distribution of energy barriers, and local structural motifs is now computationally tractable given the dramatically reduced cost of energy evaluation with MLIPs. We attributed the appearance of nearly icosahedral structures to their nearly spherical Voronoi volumes. However, at low core volume fractions, the particles are also expected to develop effective attractions to relieve entropic frustration. Disentangling the role of Voronoi geometry from that of effective attraction remains an open question, which could be addressed by constructing a reference Lennard-Jones fluid with the same activation energy to remove the geometric contribution. \par
A particularly striking finding is that the GNN predicted equilibrium energies with DFT-level accuracy and recovered equilibrium structures in agreement with experiments, despite being trained only on out-of-equilibrium data with high-energy configurations. This robust out-of-distribution generalization implies the existence of hidden structural features, possibly shared between equilibrium and out-of-equilibrium structures, that the network extracts during training. These structural features extend beyond pair distribution functions and even local bond order parameters studied here. Identifying and interpreting these hidden features could motivate new physics-based order parameters and deepen our understanding of the structure–dynamics relationship in glasses.\cite{jung2025roadmap} More broadly, this result suggests a practical workflow: train on a diverse set of accessible configurations, such as high-temperature trajectories or unbiased random snapshots, and deploy at conditions that would otherwise be computationally prohibitive. Whether this generalization extends to other glassy systems remains to be tested.  \par 
Our results demonstrate that recent advances in MLIPs can be readily translated to many-body colloidal and soft-matter systems. Prior machine learning approaches in these systems have largely been restricted to low-dimensional training data involving only two- and three-body potentials of mean force.\cite{campos2021machine,zhou2023many,giunta2023coarse,campos2024machine,alkemade2025machine,zhang2025generalized} These approaches typically rely on invariant, hand-crafted descriptors that demand large training sets and generalize poorly to unseen conditions. Furthermore, the exceptional accuracy and data-efficiency of equivariant models show great promise in inverse design of amorphous materials with targeted properties and structures, particularly when many-body interactions cannot be neglected. Such frameworks can be further extended to enhance the understanding and control of colloidal self-assembly\cite{dijkstra2021predictive} and to design polymer nanocomposites with controlled properties.\cite{chremos2016self, hansoge2018materials}

\section*{Methods} \label{sec:methods}
\subsection*{Classical DFT} \label{subsec:classical DFT}
The potential energy of a given particle configuration, $E(\mathbf{R}_1, \dots, \mathbf{R}_N)$, normalized by the Boltzmann constant, $k_\mathrm{B}$, and temperature, $T$, can be written as follows:
\vspace{-3pt}
\begin{equation} \label{Eq:many-body-potential}
\begin{split}
\frac{E(\mathbf{R}_1, \dots, \mathbf{R}_N)}{k_\mathrm{B}T} = &\bigintsss_V \bigintsss_A \left[P(\mathbf{r},\mathbf{r_0}) \left[\ln \left(P(\mathbf{r},\mathbf{r_0})\Lambda_b^3  \right) - 1 \right] + P(\mathbf{r},\mathbf{r_0})\frac{|\mathbf{r}-\mathbf{r}_0|^2}{4R_g^2}\right] d\mathbf{r}_0 \, d\mathbf{r} \\
&+ \frac{\alpha}{2}\bigintsss_V \, \left(n(\mathbf{r})-n_0 \right)^2 d\mathbf{r}.
\end{split}
\end{equation}
We model the polymers as bead-springs attached to the surface of the cores, where $P(\mathbf{r},\mathbf{r}_0)$ represents the probability density of finding a polymer bead position $\mathbf{r}$ given that it is grafted at a position $\mathbf{r}_0$. Within the first integral, the first term captures the translational entropy of the polymers via the ideal gas Helmholtz free energy of the beads, where $\Lambda_b$ is the thermal de Broglie wavelength of the beads, $V$ is the volume of the system, and $A$ is the surface area of all the cores. The second term models the configurational entropy of polymer chains by linear and massless springs with energy $E_{\mathrm{spring}} = \frac{1}{2} \xi x^2$ where $\xi = \frac{k_B T}{2R_g^2}$ is the spring constant, $x$ is the displacement of the spring from its rest length of zero, and $R_g$ is the radius of gyration of an ideal, unattached linear chain. The last term imposes the incompressibility constraint as a free energy penalty based on an incompressibility parameter, $\alpha$. In the limit of $\alpha \rightarrow \infty$, there is a large free energy penalty for any number density, $n(\mathbf{r})$, deviations from the mean value, $n_0$. This corresponds to the solvent-free case, whereas $\alpha=0$ can be viewed as having excess implicit, theta solvent. The specific form of the energy penalty is derived from the variations in free energy of a Lennard-Jones liquid of monomer beads, where $\alpha$ is a function of the ratio of the inter-bead potential to the thermal energy.\cite{chremos2011structure,betancourt2008equation} The physical implications of different $\alpha$ values are explained in our previous work.\cite{ghomsheh2026linking} Here, we mainly focus on the incompressible solvent-free limit ($\alpha \rightarrow \infty$). The incompressibility constraint effectively replaces excluded volume considerations of monomers in solvent-swollen polymer brushes.\cite{de1980conformations} The excluded volume interactions between the cores and the polymers were enforced by modeling the core impenetrability via a soft potential which makes the integrands vanish inside the cores. \par 
% For intermediate values of $\alpha$, the energy penalty models the finite compressibility of grafted polymers in a solvent-free system.
For a given configuration of cores, the probability density of the grafted polymers at equilibrium is obtained by minimizing their free energy:
\begin{equation} \label{Eq:polymer-probability-density}
    \frac{\delta E(\mathbf{R}_1, \dots, \mathbf{R}_N)}{\delta P} = 0 \rightarrow P(\mathbf{r},\mathbf{r}_0) = K(\mathbf{r}_0)\exp{\left[-\frac{|\mathbf{r}-\mathbf{r}_0|^2}{4R_g^2}\right]} \exp{[-\alpha(n(\mathbf{r})-n_0)]}
\end{equation}
where $K(\mathbf{r}_0)$ is a normalization coefficient determined by the grafting density, $\sigma_g$, as follows:
\begin{equation} \label{Eq:polymer-grafting-density}
	\sigma_g = \int_V \, P(\mathbf{r},\mathbf{r}_0) \, d\mathbf{r} = \frac{N_g}{A_c}.
\end{equation}
Here, $N_g$ is the number of grafted polymers per particle, and $A_c$ is the surface area of a single core. The mean number density is defined as $n_0 = \frac{NN_g}{V_{\text{void}}}$, where $V_{\text{void}}$ represents the void volume in the system. The number density, $n(\mathbf{r})$, of the grafted polymers is given by:
\begin{equation} \label{Eq:polymer-number-density}
		n(\mathbf{r}) = \int_A \, P(\mathbf{r},\mathbf{r_0}) \, d\mathbf{r_0}.
\end{equation}
We simulated $N=100$ monodisperse, solvent-free PGNs and set the core diameter, $d$, as the unit of length. To calculate the potential energy, $E(\mathbf{R}_1, \dots, \mathbf{R}_N)$, for each particle configuration, we must determine the probability and number density of the polymers at every point in the space. This requires solving a highly coupled system of equations (Eqs. \eqref{Eq:polymer-probability-density} to \eqref{Eq:polymer-number-density}). Naively, these coupled volume and area integrals lead to a quadratic computational scaling with the number of particles. We introduced local neighborhood cutoffs around each particle. Restricting the volume integrals exclusively to these local regions results in linear scaling with the number of particles. The size of these local regions was chosen to be sufficiently larger than typical inter-particle distances such that polymer stretches beyond this distance carry negligible probability. This cutoff only constrains the reach of individual chains and does not restrict the range of many-body interactions. To solve for $\alpha \rightarrow \infty$, we employed an iterative scheme detailed in our previous work,\cite{ghomsheh2026linking} gradually increasing $\alpha$ while using the previous solution as an initial guess. 
% Numerical integration was performed using a fourth-order accurate, six-point quadrature rule\cite{hughes2003finite} for the volume integrals and Lebedev quadrature rule\cite{ericson2001codes} for the surface integrals over the cores.

\subsection*{Hard-sphere configuration sampling of the PEL} \label{subsec:random_sampling}
We randomly sampled the configurational space of $N$ nanoparticles at each core volume fraction using the LAMMPS package. Particles were placed at random positions, and initial overlaps were eliminated by applying a damped harmonic repulsive potential. The system was then evolved under hard-sphere-like dynamics at finite temperature, with a harmonic spring potential preventing particle contact. Snapshots of particle positions were collected at large regular intervals such that the mean squared displacement (MSD) of the particles in successive snapshots exceeded the square of the simulation box size. The potential energy of PGNs with each of these core configurations was then calculated with our classical DFT, producing a dataset of configuration–energy pairs used to train the GNN. Since configurations are sampled without energy bias, the dataset spans a broad and diverse region of the PEL, including high-energy configurations that would be inaccessible in equilibrium sampling.

\subsection*{NequIP architecture} \label{subsec:nequip}
In NequIP, nodes within a cutoff radius, $r_{\mathrm{cut}}$, are connected by an edge. Node features are updated via message-passing as a tensor product of learnable radial functions (parameterized by a multilayer perceptron) and spherical harmonics to impose equivariance. The maximum degree of spherical harmonics, $l_{\mathrm{max}}$, sets the angular resolution.\cite{batzner2023} Isotropic distance-dependent potentials are well described by $l_{\mathrm{max}}=0$ (invariant representation), whereas strongly anisotropic interactions require higher $l_{\mathrm{max}}$. Many-body correlations are captured through $n_l$ successive message-passing layers, leading to chain-like propagation of information over an effective receptive field of $n_l \, r_{\mathrm{cut}}$. A single message-passing layer captures only pairwise interactions while each additional layer raises the correlation order by one. However, this breaks down when using nonlinear activations and readout as such nonlinearities generate implicit higher-order terms.\cite{batatia2025design} \par 
Hyperparameters were optimized using an extensive sweep monitored via Weights \& Biases.\cite{wandb} We used NequIP version 0.13.0 and trained on $4,000$ configurations with $500$ held out for validation and $500$ for testing. For learning curves, we fixed the validation and test sets and only added new frames to the existing training dataset. For all samples, the optimal angular resolution and number of message-passing layers are $l_{\mathrm{max}}=3$ and $n_l=3$, respectively. The optimal cutoff radius scales with inter-particle spacing, $r_{\mathrm{cut}} \thicksim \phi_c^{-1/3}$, and is equal to $r_{\mathrm{cut}}/d = 6$ for $\phi_c=\num{0.1}$. All remaining hyperparameters were set to the recommended defaults in \hyperlink{https://github.com/mir-group/nequip}{https://github.com/mir-group/nequip}. In Figs. \ref{fig4b}–\ref{fig4d}, the hyperparameter under investigation is varied while all others are held at their optimal values. Further details of the equivariant architecture and NequIP hyperparameters can be found in References\cite{batzner2023, batatia2025design,geiger2022e3nn,kozinsky2023scaling,tan2025high}. All models were trained using single-GPU training on an NVIDIA A100 GPU with float32 precision. 

\subsection*{Monte Carlo simulations} \label{subsec:mc}
We performed Markov Chain Monte Carlo (MCMC) sampling in the canonical ensemble. At each step, a particle is selected at random and displaced by a random vector. The energy of the proposed configuration was obtained by the GNN, and the moves were accepted or rejected according to Metropolis criterion. Moves leading to core overlaps were rejected immediately. The maximum displacement magnitude in each Cartesian direction was tuned to achieve an acceptance ratio of around $30\%$. To verify equilibration, we started the simulations from four uncorrelated disordered configurations and a FCC configuration, and monitored the evolution of the energy and structure. All runs converged to the same structure with a mean per-particle energy difference of less than $\qty{}{\kBT}$ between any two replicas. Systems at lower $\phi_c$ required more MC cycles than those at higher $\phi_c$, consistent with the deeper energy barriers at those conditions. After equilibration, configurations were collected every 20,000 MC moves, an interval confirmed to exceed the energy autocorrelation time, yielding 2,000 uncorrelated snapshots per sample for analysis. All calculations were performed on an NVIDIA A100 GPU.

\subsection*{Structure analysis} \label{subsec:structure}
We analyzed the structure using the freud library.\cite{freud2020} For each particle $i$, the complex bond-order vector is defined as\cite{steinhardt1983bond}:
\begin{equation} \label{Eq:steinhardt-qlm}
		q_{lm} (i) = \frac{1}{N_b (i)} \sum_{j=1}^{N_b(i)} Y_{lm}(\hat{\mathbf{r}}_{ij})
\end{equation}
where $N_b(i)$ is the number of nearest neighbors of particle $i$, $Y_{lm}(\textbf{r}_{ij})$ are spherical harmonics of degree $l$ and order $m=-l, \dots, l$, evaluated along the unit bond vector $\hat{\mathbf{r}}_{ij}$. The rotationally invariant Steinhardt order parameters are defined as\cite{steinhardt1983bond}: 
\begin{equation} \label{Eq:steinhardt-ql}
q_l = \sqrt{\frac{4\pi}{2l+1} \sum_{m=-l}^{l} |q_{l m}|^2},
\end{equation}

\begin{equation} \label{Eq:steinhardt-wl}
w_l = \sum_{m_1+m_2+m_3=0} \begin{pmatrix} l & l & l \\ m_1 & m_2 & m_3 \end{pmatrix} q_{l m_1} q_{l m_2} q_{l m_3},
\end{equation}

\begin{equation} \label{Eq:steinhardt-whatl}
\hat{w}_l = w_l \left( \sum_{m=-l}^{l} |q_{l m}|^2 \right)^{-\frac{3}{2}},
\end{equation}
where the bracketed symbol in Eq.~\eqref{Eq:steinhardt-wl} is the Wigner $3j$ symbol. To improve sensitivity to crystalline ordering, we use the neighbor-averaged order parameters that incorporate second-shell information:\cite{lechner2008accurate}
\begin{equation} \label{Eq:steinhardt-qlm-avg}
\bar{q}_{l m}(i) = \frac{1}{N_b(i) + 1} \left( q_{l m}(i) + \sum_{j=0}^{N_b(i)} q_{l m}(j) \right),
\end{equation}
and $\bar{q}_l(i)$ is obtained analogously to Eq.~\eqref{Eq:steinhardt-ql}. The pair $(\bar{q}_4, \bar{q}_6)$ discriminates between FCC, HCP, and BCC crystal structures, while $w_6$ and its normalized form $\hat{w}_6$ are sensitive indicators of local icosahedral-like order.\cite{leocmach2012roles} Nearest neighbors are identified as particles that share a Voronoi face with particle $i$. \par 

\section*{Acknowledgements}
This work used Delta GPU at NCSA through allocation
PHY250140 from the Advanced Cyberinfrastructure Coordination
Ecosystem: Services \& Support (ACCESS) program, which is supported by
U.S. National Science Foundation grants \#2138259, \#2138286, \#2138307,
\#2137603, and \#2138296.\cite{boerner2023access}

% ==========================================
% SUPPLEMENTARY INFORMATION
% ==========================================
\clearpage % Forces a new page
\section*{Supplementary Information} \label{sec:supplementary}

% Reset the counters for figures, tables, and equations
\setcounter{figure}{0}
\setcounter{table}{0}
\setcounter{equation}{0}

\renewcommand{\figurename}{Supplementary Figure}
\renewcommand{\tablename}{Supplementary Table}
\renewcommand{\theequation}{S\arabic{equation}}

\subsection*{PGN design parameters} \label{supp:PGN_parameters}
We assume that the space is filled either by the cores or the grafted polymers. Therefore, the core volume fraction, polymer molecular weight, and grafting density are related through the space-filling constraint. For a given core volume fraction, $\phi_c$, polymer molecular weight, $M_w$, and polymer mass density, $\rho_p$, the number of grafted polymers per particle, $N_g$, is given by:
\begin{equation} \label{Eq: space-filling}
		N_g = \frac{(1-\phi_c)V_{\mathrm{core}} \,\rho_p N_A}{\phi_c \, M_w}
\end{equation}
where $ V_{\mathrm{core}}$ is the volume of a single core, $N_A$ is Avogadro's number, $\frac{1-\phi_c}{\phi_c} V_{\mathrm{core}}$ represents the void volume per core, and $\frac{M_w}{\rho_p N_A}$ represents the volume per polymer molecule. Consistent with previous experiments,\cite{liu2019microscopic,choudhury2015self,agrawal2016molecular,srivastava2017self} we choose polyethylene glycol (PEG) as the grafted polymer. Table \ref{Table1-supp} lists the design parameters of solvent-free PGNs chosen in this study.

\begin{table}[!ht]
    \centering
    \caption{Solvent-free PGN design parameters}
    \begin{tabular}{lc}
        \toprule
        Parameter & Value \\
        \midrule
        Core particle diameter, $d$ & $\qty{10}{\nano\meter}$ \\
        Core particle volume fraction & $\num{0.1} - \num{0.4}$ \\
        Polymer molecular weight, $M_w$ & $\num{5}$ and $\qty{9}{\kilo\dalton}$ \\
        Ratio of polymer radius of gyration to particle diameter, $R_g/d$ & $\num{0.27}$ and $\num{0.37}$ \\
        Polymer mass density, $\rho_p$ & $\qty{1}{\gram\per\centi\meter\cubed}$ \\
        Grafting density, $\sigma_g$ & $\num{0.3} - \qty{1.8}{\chain\per\nano\meter\squared}$ \\
        \bottomrule
    \end{tabular}
    \label{Table1-supp}
\end{table}

\begin{figure} [!ht]
    \centering
    \begin{subfigure}{0.48\textwidth}
        \centering
        \includegraphics[width=\textwidth]{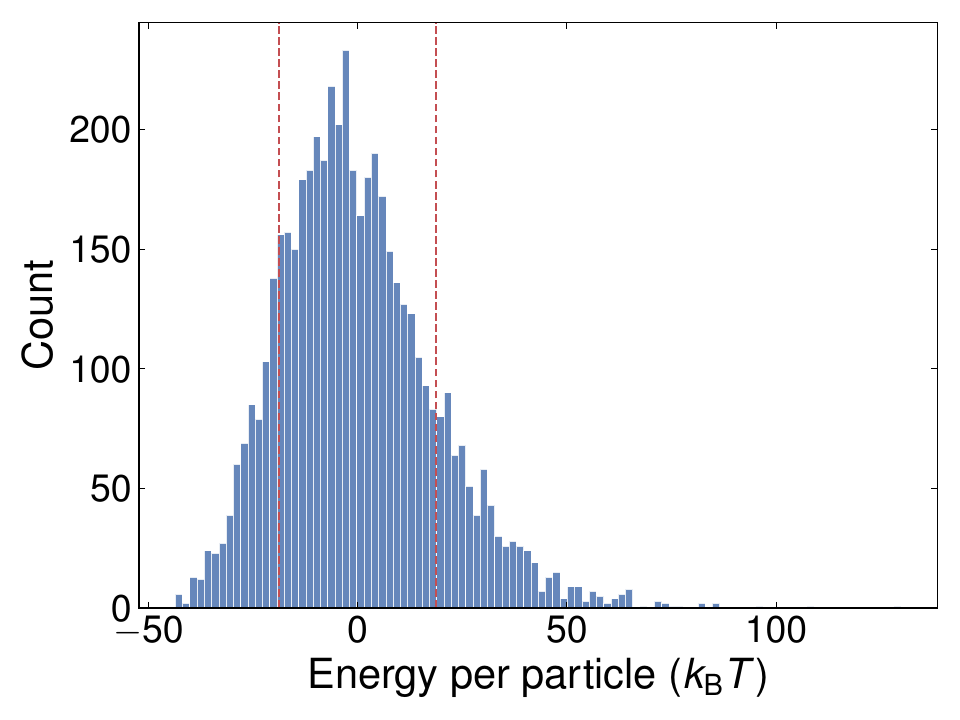}
        \caption{}
        \label{sup-fig-energydist-a}
    \end{subfigure}
    \hfil
    \begin{subfigure}{0.48\textwidth}
        \centering
        \includegraphics[width=\textwidth]{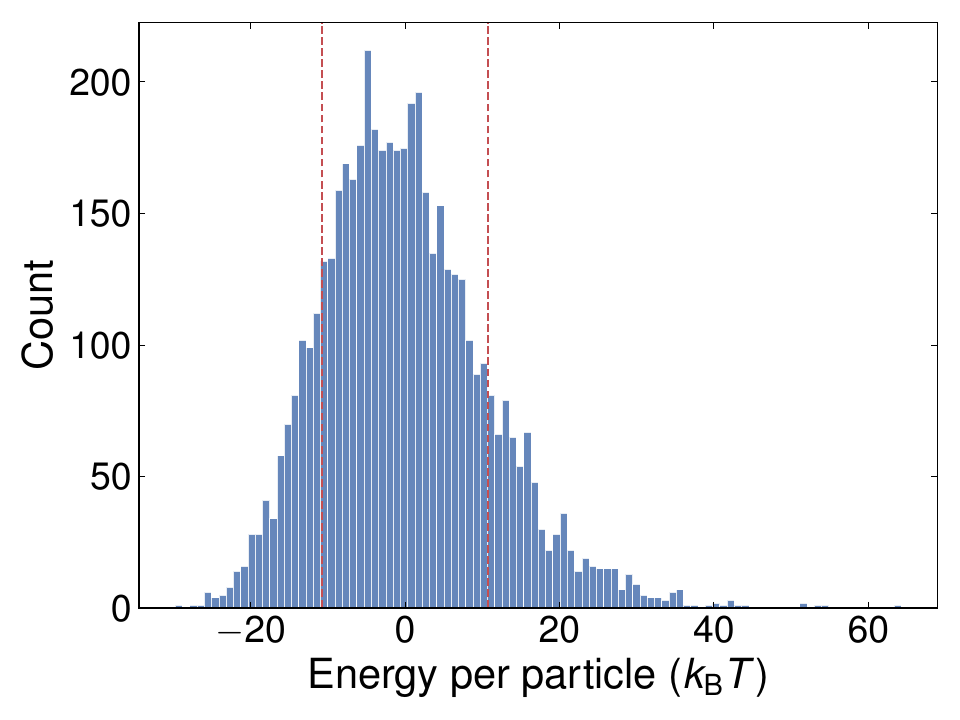}
        \caption{}
        \label{sup-fig-energydist-b}
    \end{subfigure}
    \hfil
    \begin{subfigure}{0.48\textwidth}
        \centering
        \includegraphics[width=\textwidth]{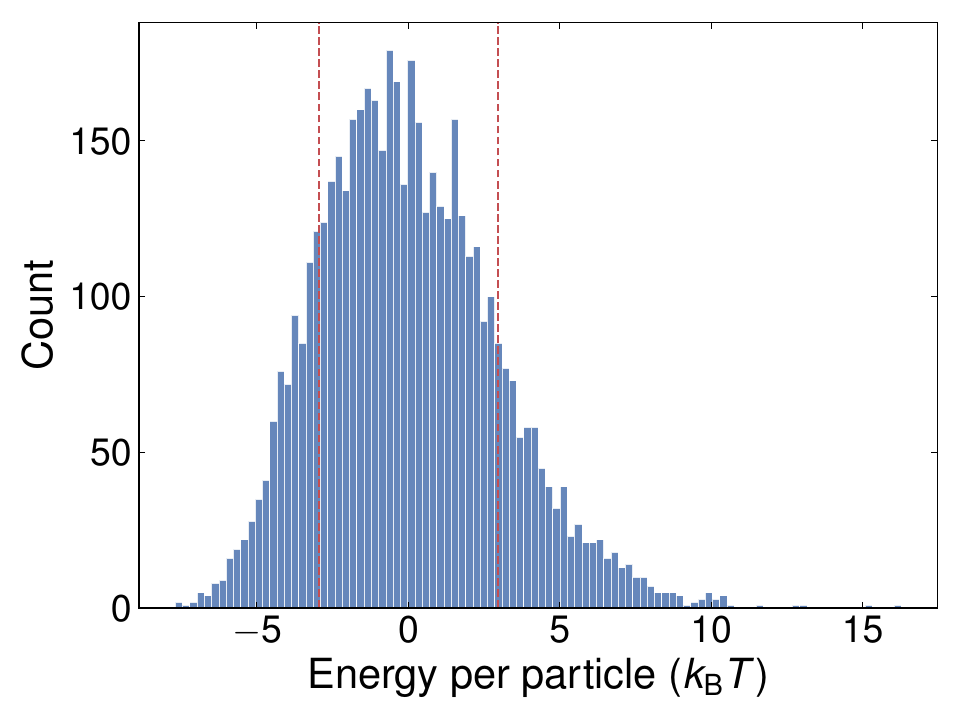}
        \caption{}
        \label{sup-fig-energydist-c}
    \end{subfigure}
    \hfil
    \begin{subfigure}{0.48\textwidth}
        \centering
        \includegraphics[width=\textwidth]{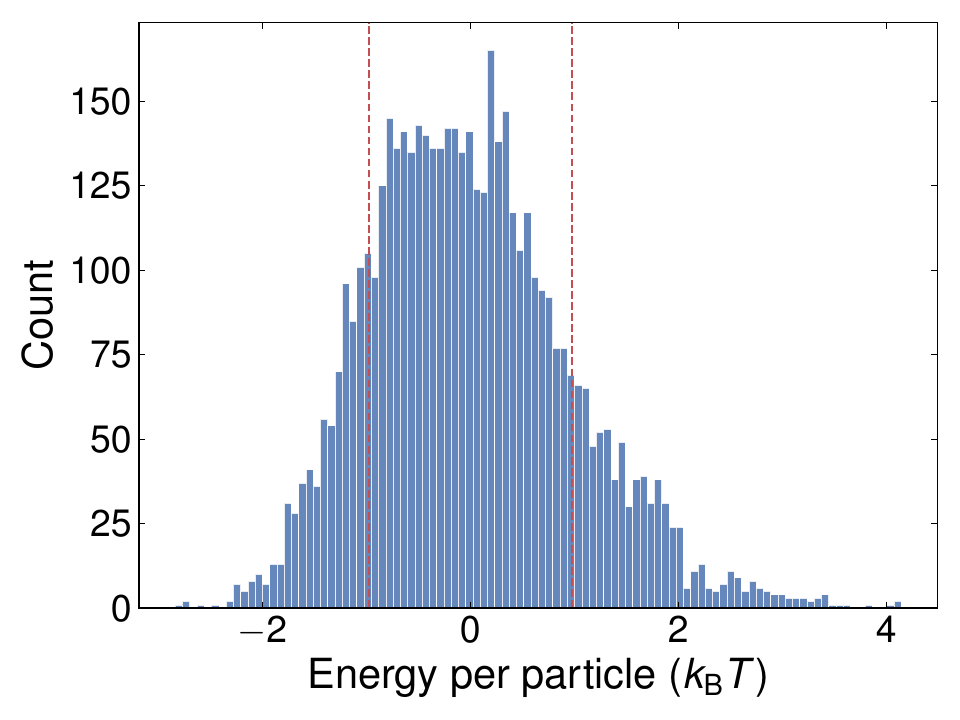}
        \caption{}
        \label{sup-fig-energydist-d}
    \end{subfigure}
    \caption{Per particle energy distribution of out-of-equilibrium configurations, shifted by the mean, for solvent-free PGNs with (a) $\phi_c = \num{0.1}$, $\sigma_g = \qty{1}{\chain\per\nano\meter\squared}$, and $M_w = \qty{9}{\kilo\dalton}$, (b) $\phi_c = \num{0.2}$, $\sigma_g = \qty{0.8}{\chain\per\nano\meter\squared}$, and $M_w = \qty{5}{\kilo\dalton}$, (c) $\phi_c = \num{0.3}$, $\sigma_g = \qty{0.5}{\chain\per\nano\meter\squared}$, and $M_w = \qty{5}{\kilo\dalton}$, and (d) $\phi_c = \num{0.4}$, $\sigma_g = \qty{0.3}{\chain\per\nano\meter\squared}$, and $M_w = \qty{5}{\kilo\dalton}$. The red vertical lines represent the standard deviation.}
    \label{sup-fig-energydist}
\end{figure}

\begin{figure} [!ht]
    \centering
    \begin{subfigure}{0.48\textwidth}
        \centering
        \includegraphics[width=\textwidth]{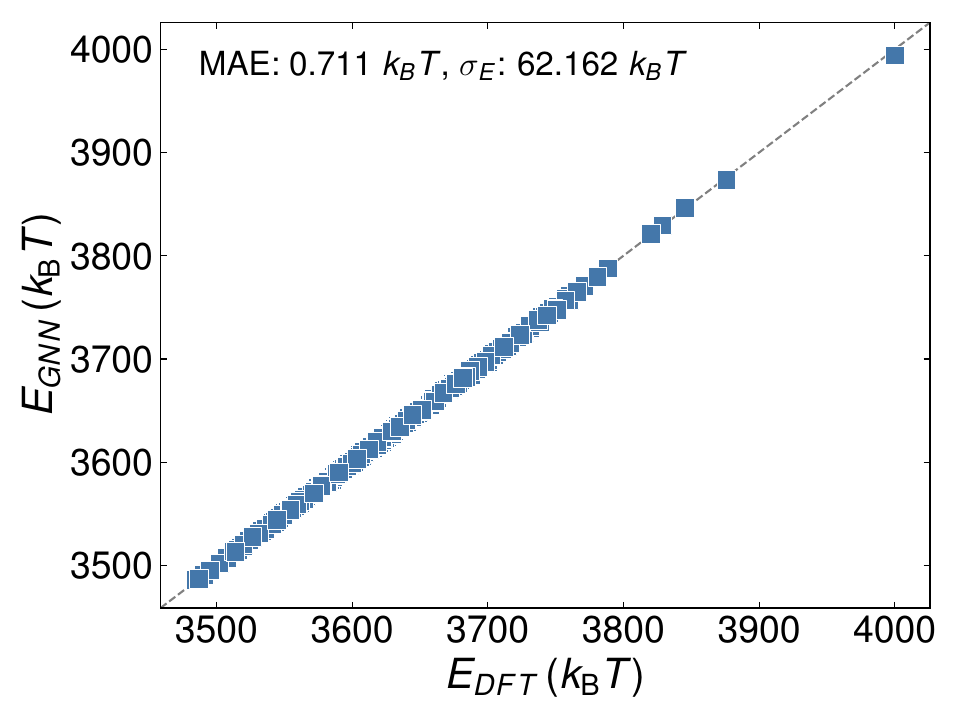}
        \caption{}
        \label{sup-fig-gnn-a}
    \end{subfigure}
    \hfil
    \begin{subfigure}{0.48\textwidth}
        \centering
        \includegraphics[width=\textwidth]{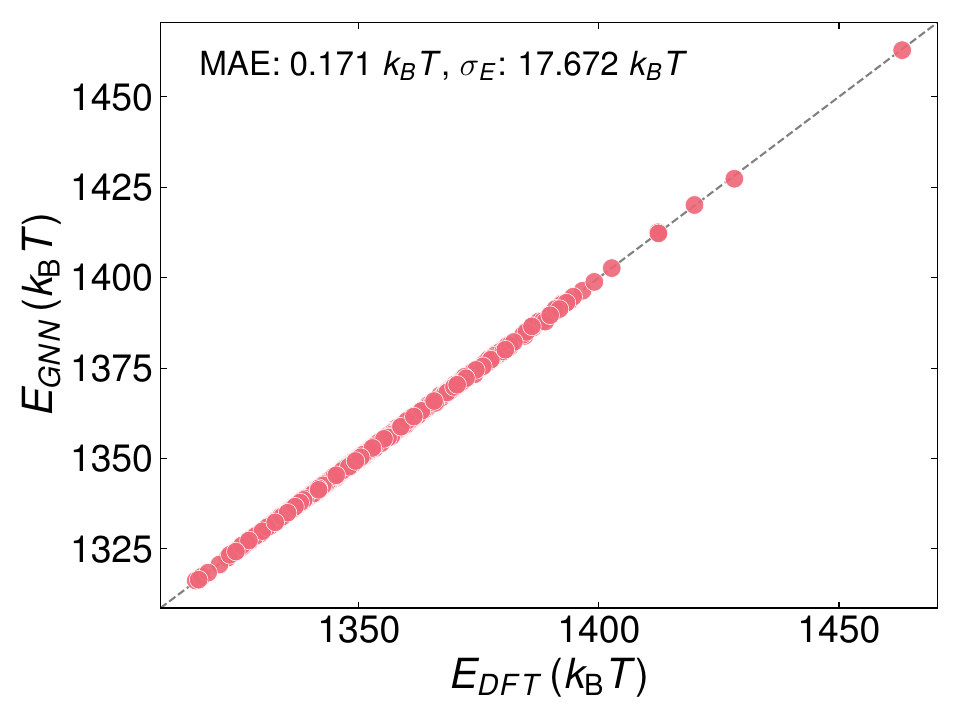}
        \caption{}
        \label{sup-fig-gnn-b}
    \end{subfigure}
    \hfil
    \begin{subfigure}{0.48\textwidth}
        \centering
        \includegraphics[width=\textwidth]{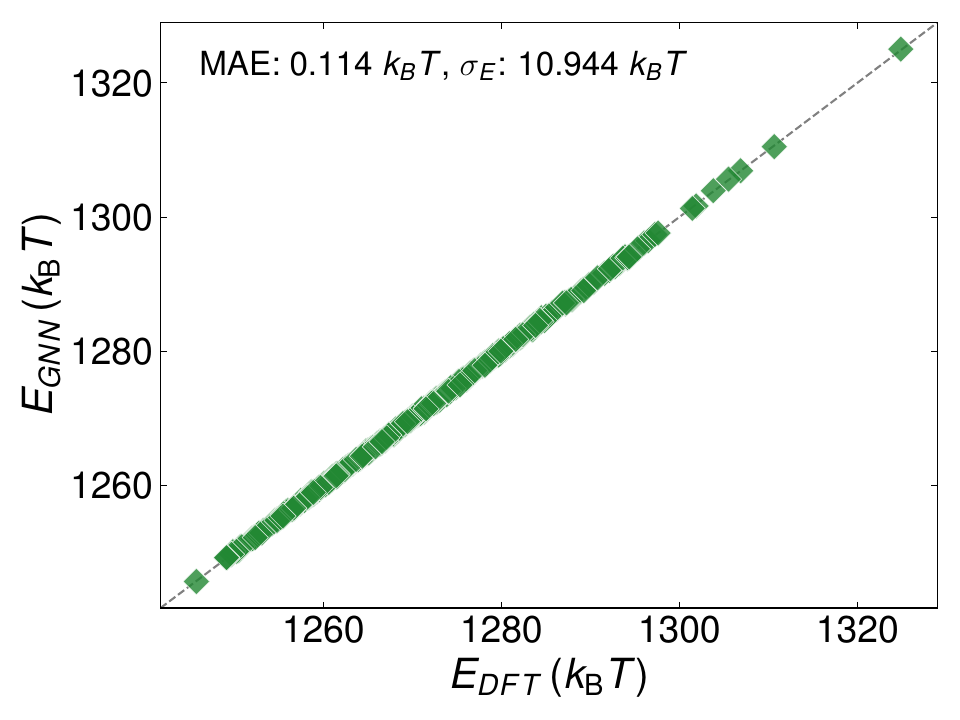}
        \caption{}
        \label{sup-fig-gnn-c}
    \end{subfigure}
    \hfil
    \begin{subfigure}{0.48\textwidth}
        \centering
        \includegraphics[width=\textwidth]{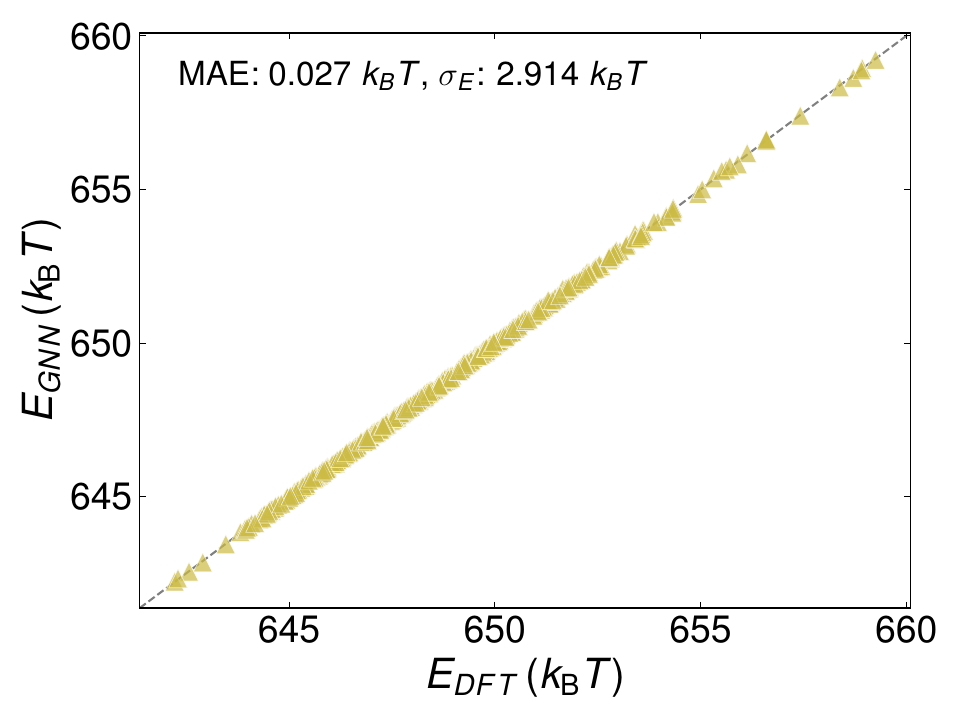}
        \caption{}
        \label{sup-fig-gnn-d}
    \end{subfigure}
    \hfil
    \begin{subfigure}{0.48\textwidth}
        \centering
        \includegraphics[width=\textwidth]{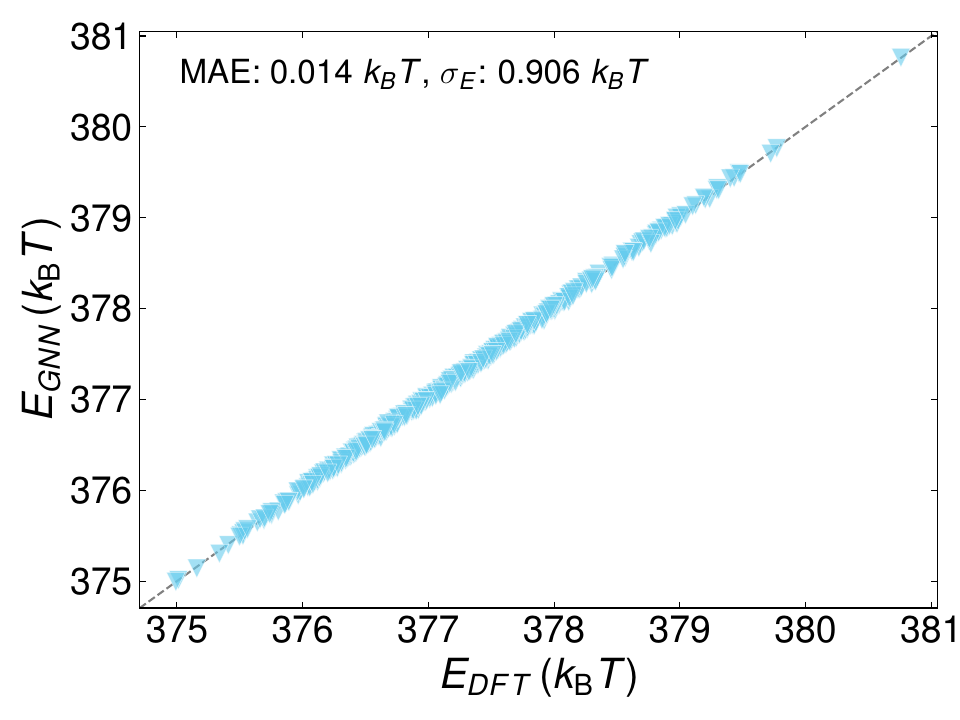}
        \caption{}
        \label{sup-fig-gnn-e}
    \end{subfigure}
    \caption{Parity plots comparing per-particle energies of out-of-equilibrium configurations predicted by classical DFT and GNN for solvent-free PGNs with (a) $\phi_c = \num{0.1}$, $\sigma_g = \qty{1.8}{\chain\per\nano\meter\squared}$, and $M_w = \qty{5}{\kilo\dalton}$, (b) $\phi_c = \num{0.1}$, $\sigma_g = \qty{1}{\chain\per\nano\meter\squared}$, and $M_w = \qty{9}{\kilo\dalton}$, (c) $\phi_c = \num{0.2}$, $\sigma_g = \qty{0.8}{\chain\per\nano\meter\squared}$, and $M_w = \qty{5}{\kilo\dalton}$, (d) $\phi_c = \num{0.3}$, $\sigma_g = \qty{0.5}{\chain\per\nano\meter\squared}$, and $M_w = \qty{5}{\kilo\dalton}$, and (e) $\phi_c = \num{0.4}$, $\sigma_g = \qty{0.3}{\chain\per\nano\meter\squared}$, and $M_w = \qty{5}{\kilo\dalton}$. The per-particle MAEs and energy standard deviations are displayed on each plot.}
    \label{sup-fig-gnn}
\end{figure}

\begin{table}[H]
    \centering
    \caption{Per-particle mean absolute error of out-of-equilibrium energies (in $\unit{\kBT}$) between classical DFT and GNN for the training, validation, and test datasets across different samples.}
    \renewcommand{\arraystretch}{1.5} 
    \begin{tabular}{l S[table-format=1.2] S[table-format=1.2] S[table-format=3.2]}
        \toprule
        {Sample} & 
        {Training} & 
        {Validation} & 
        {Test} \\
        \midrule
        $\phi_c = \num{0.1}, \sigma_g = \qty{1.8}{\chain\per\nano\meter\squared}$ & 0.379 & 0.685 & 0.711 \\
        $\phi_c = \num{0.1}, \sigma_g = \qty{1}{\chain\per\nano\meter\squared}$ & 0.093 & 0.152 & 0.171 \\
        $\phi_c = \num{0.2}, \sigma_g = \qty{0.8}{\chain\per\nano\meter\squared}$ & 0.056 & 0.113 &  0.114 \\
        $\phi_c = \num{0.3}, \sigma_g = \qty{0.5}{\chain\per\nano\meter\squared}$ & 0.017 & 0.026 &  0.027 \\
        $\phi_c = \num{0.4}, \sigma_g = \qty{0.3}{\chain\per\nano\meter\squared}$ & 0.008 & 0.015 &   0.014 \\
        \bottomrule
    \end{tabular}
    \label{sup-table-Tr-Val-Test-Error}
\end{table}

\begin{figure} [!ht]
    \centering
    \includegraphics[width=0.48\textwidth]{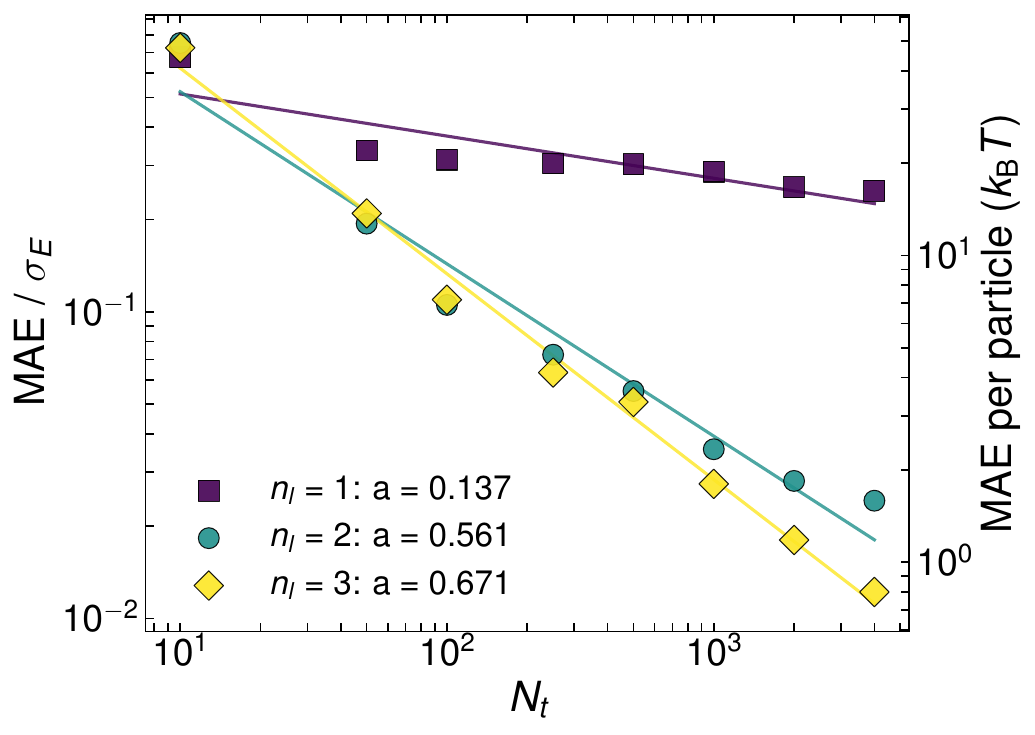}
    \caption{Learning curve for different number of message-passing layers. Nonlinear activations between message-passing layers are omitted in this test.}
    \label{sup-fig-lc-nl}
\end{figure}

\begin{figure} [!ht]
    \centering
    \begin{subfigure}{0.48\textwidth}
        \centering
        \includegraphics[width=\textwidth]{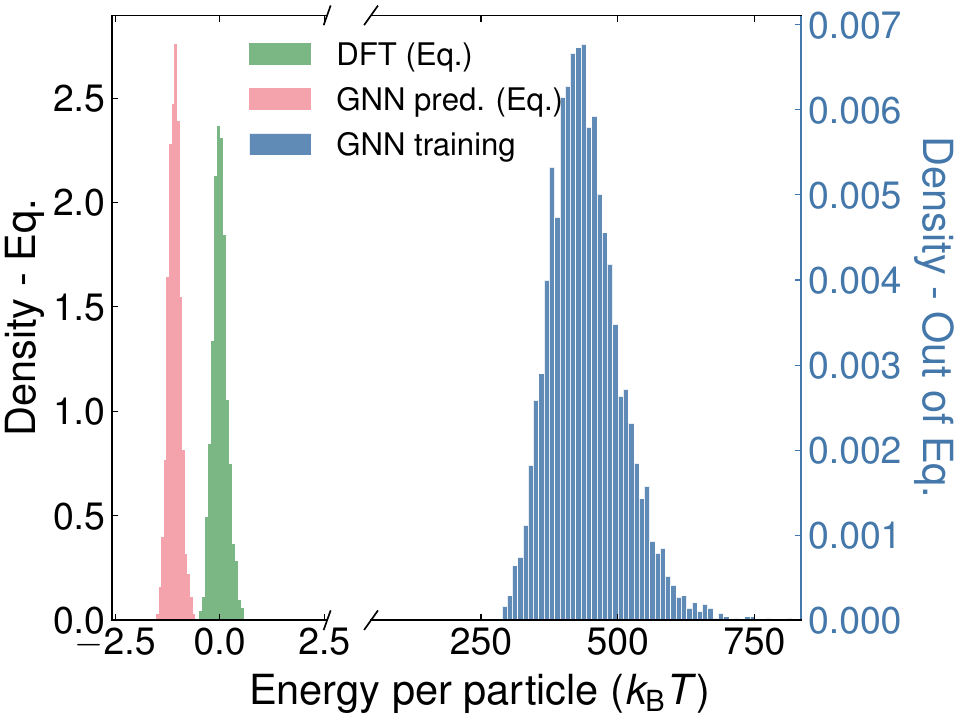}
        \caption{}
        \label{eq-en-phip1}
    \end{subfigure}
    \hfil
    \begin{subfigure}{0.48\textwidth}
        \centering
        \includegraphics[width=\textwidth]{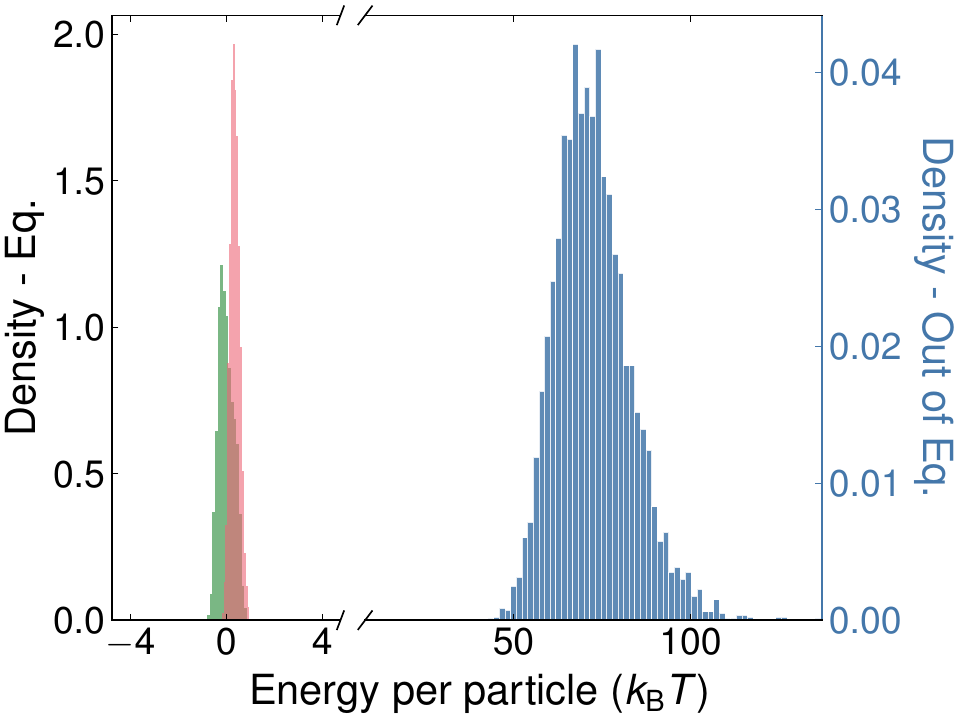}
        \caption{}
        \label{eq-en-phip2}
    \end{subfigure}
    \hfil
    \begin{subfigure}{0.48\textwidth}
        \centering
        \includegraphics[width=\textwidth]{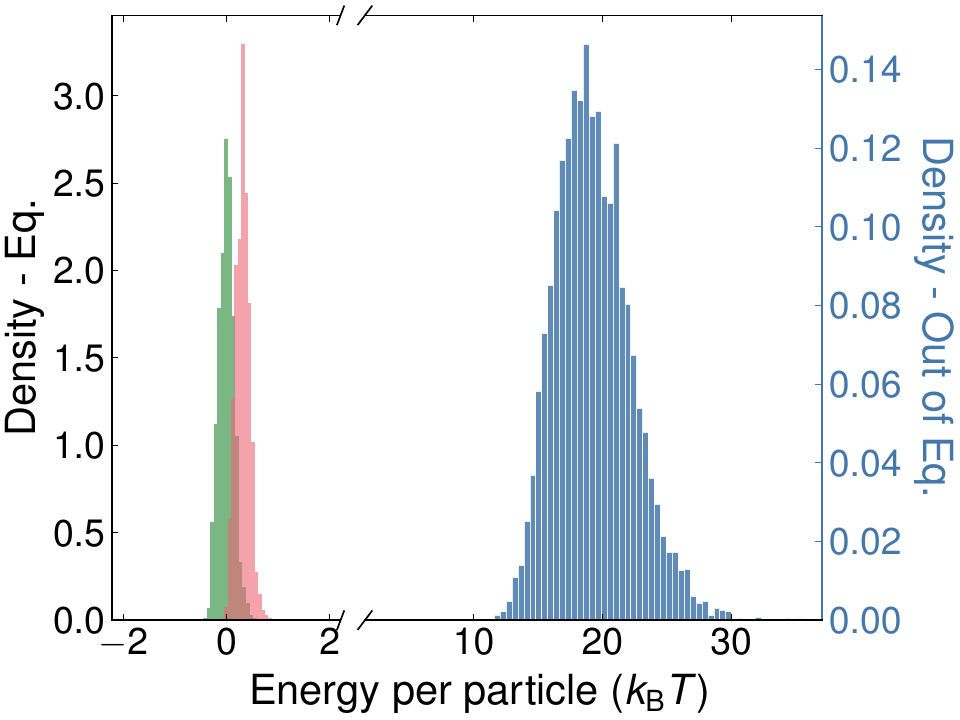}
        \caption{}
        \label{eq-en-phip3}
    \end{subfigure}
    \hfil
    \begin{subfigure}{0.48\textwidth}
        \centering
        \includegraphics[width=\textwidth]{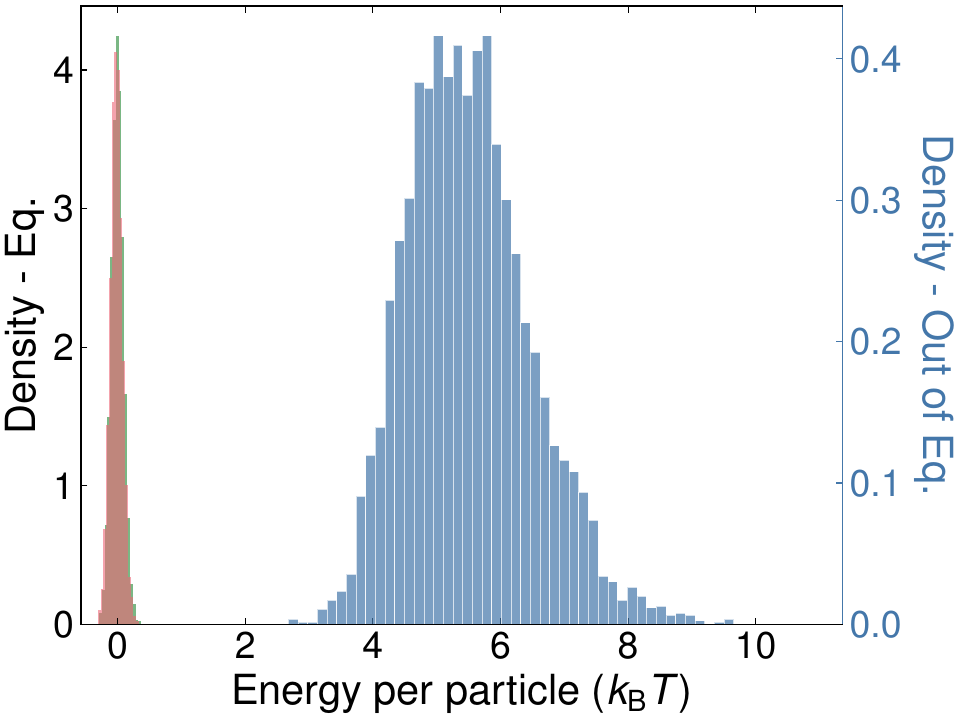}
        \caption{}
        \label{eq-en-phip4}
    \end{subfigure}
    \caption{Per particle energy distribution for out-of-equilibrium configurations (sampled by hard-sphere dynamics and used for training GNN) and equilibrium configurations evaluated by GNN and classical DFT for the sample with (a) $\phi_c = \num{0.1}$, $\sigma_g = \qty{1.8}{\chain\per\nano\meter\squared}$, and $M_w = \qty{5}{\kilo\dalton}$, (b) $\phi_c = \num{0.2}$, $\sigma_g = \qty{0.8}{\chain\per\nano\meter\squared}$, and $M_w = \qty{5}{\kilo\dalton}$, (c) $\phi_c = \num{0.3}$, $\sigma_g = \qty{0.5}{\chain\per\nano\meter\squared}$, and $M_w = \qty{5}{\kilo\dalton}$, and (d) $\phi_c = \num{0.4}$, $\sigma_g = \qty{0.3}{\chain\per\nano\meter\squared}$, and $M_w = \qty{5}{\kilo\dalton}$.}
    \label{eq-energies-dft-vs-neq}
\end{figure}

\begin{table}[H]
    \centering
    \caption{Mean absolute error of energies and relative energies between classical DFT and GNN for equilibrium configurations across samples. All quantities are per particle. The energy MAE is calculated as $\mathrm{MAE}_{\mathrm{E}} = \frac{1}{N_{\mathrm{config}}} \sum_{i=1}^{N_{\mathrm{config}}} \left| E_i^{\mathrm{DFT}} - E_i^{\mathrm{GNN}} \right|$, and the relative energy MAE is defined as $\mathrm{MAE}_{\mathrm{\Delta E}} = \frac{1}{N_{\mathrm{config}}} \sum_{i=1}^{N_{\mathrm{config}}} \left| \Delta E_i^{\mathrm{DFT}} - \Delta E_i^{\mathrm{GNN}} \right|$ where $\Delta E_i = E_i - \bar{E}$ and $\bar{E}$ is the mean energy of the equilibrium configurations, evaluated separately for classical DFT and GNN. The last column gives the separation between the mean energies of the out-of-equilibrium (training) and equilibrium configurations.}
    \renewcommand{\arraystretch}{1.5} 
    \begin{tabular}{l S[table-format=1.2] S[table-format=1.2] S[table-format=3.2]}
        \toprule
        {Sample} & 
        {$\frac{\mathrm{MAE}_{\mathrm{E}}}{N} \, (\qty{}{\kBT})$} & 
        {$\frac{\mathrm{MAE}_{\mathrm{\Delta E}}}{N} \, (\qty{}{\kBT})$} & 
        {$\frac{\bar{E}_{\mathrm{HS}} - \bar{E}_{\mathrm{Eq}}}{N} \, (\qty{}{\kBT})$} \\
        \midrule
        $\phi_c = \num{0.1}, \sigma_g = \qty{1.8}{\chain\per\nano\meter\squared}$ & 1.08 & 0.15 & 442.03 \\
        $\phi_c = \num{0.2}, \sigma_g = \qty{0.8}{\chain\per\nano\meter\squared}$ & 0.36 & 0.10 &  72.57 \\
        $\phi_c = \num{0.3}, \sigma_g = \qty{0.5}{\chain\per\nano\meter\squared}$ & 0.31 & 0.04 &  19.44 \\
        $\phi_c = \num{0.4}, \sigma_g = \qty{0.3}{\chain\per\nano\meter\squared}$ & 0.02 & 0.01 &   5.51 \\
        \bottomrule
    \end{tabular}
    \label{sup-table-DFT-GNN-Error}
\end{table}

\begin{figure} [!ht]
    \centering
    \begin{subfigure}{0.48\textwidth}
        \centering
        \includegraphics[width=\textwidth]{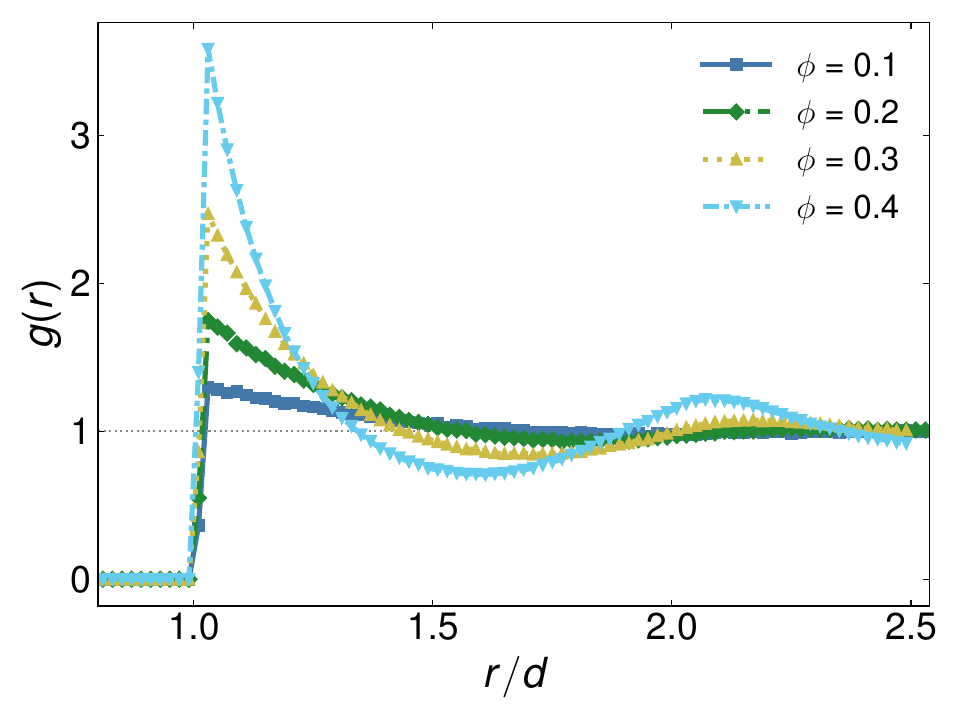}
        \caption{}
        \label{gr-hs}
    \end{subfigure}
    \hfil
    \begin{subfigure}{0.48\textwidth}
        \centering
        \includegraphics[width=\textwidth]{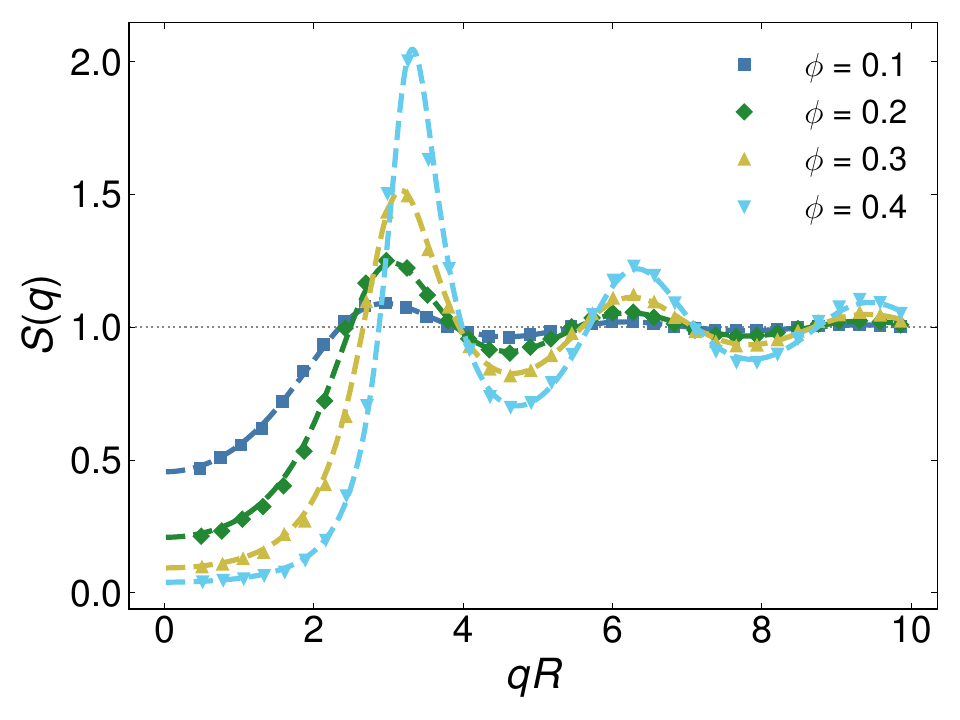}
        \caption{}
        \label{Sq-hs}
    \end{subfigure}
    \caption{(a) Pair distribution function and (b) static structure factor for out-of-equilibrium configurations sampled via hard-sphere dynamics. The dashed lines in (b) are Percus–Yevick approximation for hard-sphere liquids.}
    \label{HS-structure}
\end{figure}

\begin{figure} [!ht]
    \centering
    \begin{subfigure}{0.48\textwidth}
        \centering
        \includegraphics[width=\textwidth]{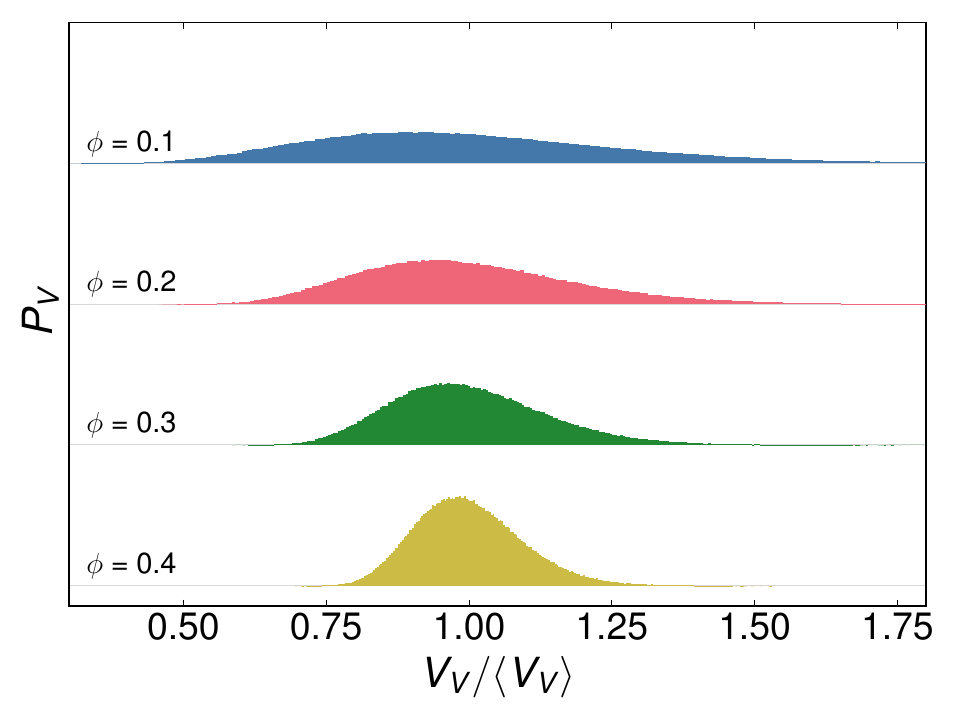}
        \caption{}
        \label{voronoi-init}
    \end{subfigure}
    \hfil
    \begin{subfigure}{0.48\textwidth}
        \centering
        \includegraphics[width=\textwidth]{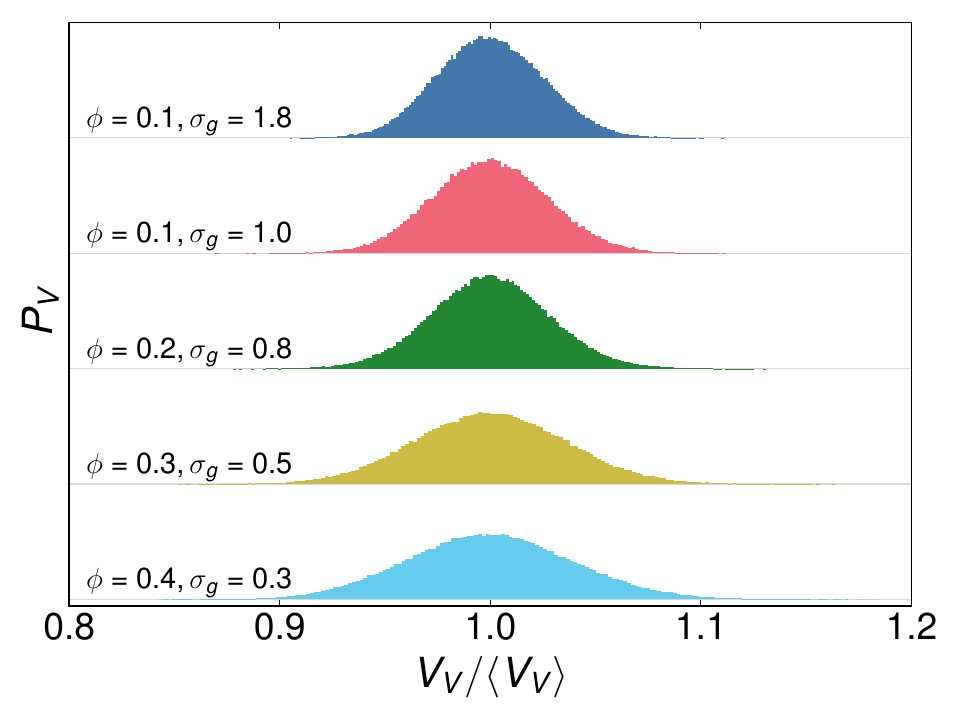}
        \caption{}
        \label{voronoi-eq}
    \end{subfigure}
    \caption{The Voronoi volume distribution normalized by the average Voronoi volume for (a) out-of-equilibrium and (b) equilibrium configurations at different PGN design parameters.}
    \label{voronoi}
\end{figure}

\begin{figure} [!ht]
    \centering
    \includegraphics[width=0.48\textwidth]{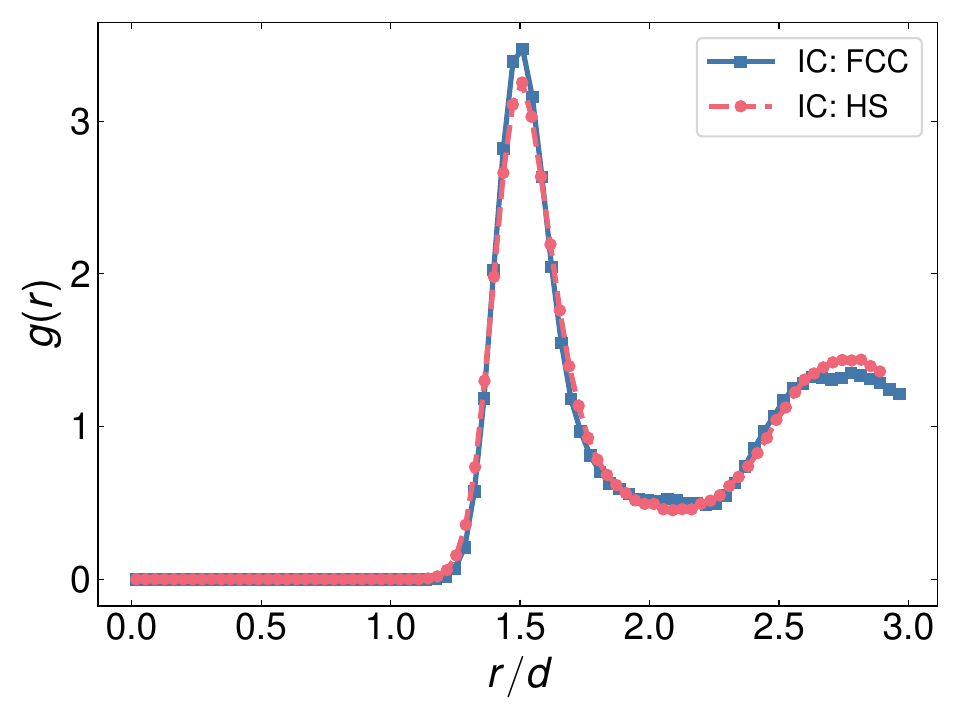}
    \caption{The equilibrium pair distribution function for solvent-free PGNs with $\phi_c=\num{0.2}$ and $\sigma_g = \qty{0.8}{\chain\per\nano\meter\squared}$ resulting from MC simulations starting from hard-sphere sampled configurations vs FCC arrangement of cores.}
    \label{fcc-vs-hs}
\end{figure}

\begin{figure} [!ht]
    \centering
    \includegraphics[width=0.48\textwidth]{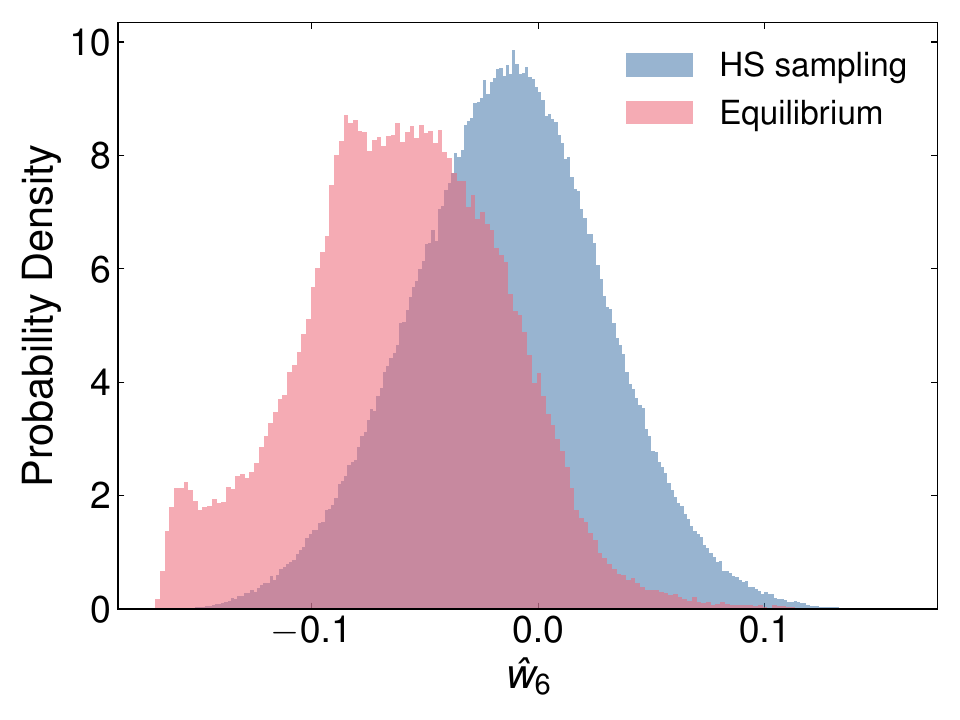}
    \caption{The Steinhardt bond order parameter $\hat{w}_6$ for solvent-free PGNs with $\phi_c=\num{0.1}$ and $\sigma_g = \qty{1.8}{\chain\per\nano\meter\squared}$ at equilibrium vs that for the structures sampled via hard-sphere dynamics at the same core volume fraction. The out-of-equilibrium structures show no local icosahedral ordering.}
    \label{w6-hs-vs-eq}
\end{figure}

% ===========================================================
% BIBLIOGRAPHY
% ===========================================================
% Use 'unsrt' for arXiv stability. 
% (For Nature submission later, you can swap this to 'naturemag' if you have the .bst file)
\clearpage
\renewcommand{\bibfont}{\small}
\bibliographystyle{unsrt}
\bibliography{references}

\end{document}